\documentclass[prd, 10pt,aps,nofootinbib, floatfix, notitlepage,twocolumn]{revtex4-1}

\usepackage{amsmath,amsfonts,amssymb}
\usepackage{mathrsfs}
\usepackage{graphicx}
\usepackage[english]{babel} 
\usepackage{hyperref} 
\usepackage{color}
\usepackage{adjustbox}

\newcommand{\be}{\begin{equation}}
\newcommand{\ee}{\end{equation}}
\newcommand{\bes}{\begin{equation*}}
\newcommand{\ees}{\end{equation*}}

  \usepackage[utf8]{inputenc}

\def\VEV#1{\left\langle #1 \right\rangle}
\newcommand{\wigner}[6]{ \begin{pmatrix}
  #1 & #2 & #3 \\
  #4 & #5 & #6 
\end{pmatrix}}

\def\wigner#1#2#3#4#5#6{ \left( \begin{array}{ccc} #1 & #3 & #5
\\ #2 & #4 & #6 \\ \end{array} \right)}

\begin{document}

\title{Probing spatial variation of the fine-structure constant using the CMB}
\author{Tristan L.~Smith$^1$}
\author{Daniel Grin$^2$}
\author{David Robinson$^1$}
\thanks{These authors contributed equally to this work.}
\author{Davy Qi$^1$} \thanks{These authors contributed equally to this work.}
\affiliation{$^1$Department of Physics and Astronomy, Swarthmore College, 500 College Ave., Swarthmore, PA 19081, United States}
\affiliation{$^2$Department of Physics and Astronomy, Haverford College, 370 Lancaster Avenue, Haverford, PA 19041, United States}

\date{\today}

\begin{abstract}
The fine-structure constant, $\alpha$, controls the strength of the electromagnetic interaction.  There are extensions of the standard model in which $\alpha$ is dynamical on cosmological length and time-scales. The physics of the cosmic microwave background (CMB) depends on the value of $\alpha$. The effects of spatial variation in $\alpha$ on the CMB are similar to those produced by weak lensing: smoothing of the power spectrum, and generation of non-Gaussian features. These would induce a bias to estimates of the weak-lensing potential power spectrum of the CMB. Using this effect, \textit{Planck} measurements of the temperature and polarization power spectrum, as well as estimates of CMB lensing, are used to place limits (95\% C.~L.) on the amplitude of a scale-invariant angular power spectrum of $\alpha$ fluctuations relative to the mean value ($C_L^{\alpha}=A^\alpha_{\rm SI}/[L(L+1)]$) of $A^\alpha_{\rm SI}\leq 1.6 \times 10^{-5}$. The limits depend on the assumed shape of the $\alpha$-fluctuation power spectrum. For example, for a white noise angular power spectrum ($C_L^{\alpha}=A^\alpha_{\rm WN}$), the limit is $A^\alpha_{\rm WN}\leq 2.3 \times 10^{-8}$. It is found that the response of the CMB to $\alpha$ fluctuations depends on a separate-universe approximation, such that theoretical predictions are only reliable for $\alpha$ multipoles with $L\lesssim 100$ . An optimal trispectrum estimator can be constructed and it is found that it is only marginally more sensitive than lensing techniques for \textit{Planck} but significantly more sensitive when considering the next generation of experiments.  For a future CMB experiment with cosmic-variance limited polarization sensitivity (e.g., CMB-S4), the optimal estimator could detect $\alpha$ fluctuations with $A^\alpha_{\rm SI}>1.9 \times 10^{-6}$ and $A^\alpha_{\rm WN} > 1.4 \times 10^{-9}$. 
\end{abstract}

\maketitle

\section{Introduction}

Ever since Paul Dirac hypothesized the Law of Large Numbers \cite{1938RSPSA.165..199D}, physicists have explored the possibility that constants of nature are not in fact constant. Dirac proposed time variation of the gravitational constant $G$ to ensure that certain large numbers in cosmology would be the same order of magnitude throughout time \cite{1938RSPSA.165..199D,Brans:1961sx}, and Gamow then suggested that time variation of the electric charge $e$ could explain the same coincidences \cite{Gamow:1967zz,Gamow:1967zza}. The time dependence required to explain these coincidences has been ruled out by stellar evolution and a variety of anthropic arguments \cite{Teller:1948zz}, but others have since explored more subtle variations in these and other fundamental constants, which emerge as predictions of theories with large extra dimensions \cite{PhysRevD.21.2167,Kolb:1985sj,PhysRevD.60.116004}. 

There are several theories which naturally incorporate a dynamical fine-structure constant, $\alpha$. Bekenstein proposed a model for a varying $\alpha$ which suppresses violations of the Weak Equivalence Principle (WEP) to undetectable levels \cite{Bekenstein:1982eu}.  The full theory, known as the Bekenstein-Sandvik-Barrow-Magueijo (BSBM) model, places Bekenstein's scalar field in a cosmological context, allowing it and $\alpha$ to evolve with the expansion of the universe. The BSBM model makes predictions for how $\alpha$ will vary in time and space \cite{Barrow:2002db,Barrow:2002zh}.  Variations and extensions of the BSBM model exist that consider other effects such as density inhomogeneities \cite{Mota:2003tm}, as well as more complicated scalar field couplings and potentials \cite{Barrow:2013uza, Graham:2014hva}, including a quintessence field \cite{Copeland:2003cv}, among others.  

There has also been growing interest in models that `disformally' couple electromagnetism to a scalar field \cite{Bekenstein:1992pj,vandeBruck:2015rma}, as well as string-inspired `runaway dilaton' models with dynamical extra dimensions that are stabilized by matter couplings in a way that yields potentially observable time evolution and spatial fluctuations in $\alpha$ \cite{Damour:1994zq,Damour:2002nv,Damour:2002mi,Martins:2017yxk}, as well as models in which a light scalar dark matter component induces $\alpha$ fluctuations \cite{Sigurdson:2003pd,Stadnik:2015kia}.

On the observational side, claims have been made that the absorption spectra of distant quasars support cosmological time variation and a dipole in $\alpha$ \cite{Webb:2010hc, King:2012id, Webb:1998cq, Murphy:2002jx, Murphy:2003hw}.  More recent observations and analyses have failed to reproduce such a result consistently \cite{Songaila:2014fza, Martins:2017qxd}, calling into question the method (in particular, the spatial stability of the wavelength calibration) used to obtain the spatial dipole result \cite{Murphy:2017xaz}. Future efforts at the Very Large Telescope (VLT) \cite{Martins:2017yxk} could improve sensitivity to $\alpha$ variations by an additional two orders of magnitude.

Other observational techniques have been used to attempt to constrain the magnitude of time variation in $\alpha$. For example, the rare-earth element abundance data from Oklo (a  naturally occurring uranium fission reactor from approximately 2 billion years ago in Gabon), which is completely independent of cosmological models, places constraints on the possible temporal variations of $\alpha$ to $-6.7\times10^{-17}\text{yr}^{-1}<\dot{\alpha}/{\alpha}<5.0\times10^{-17}\text{yr}^{-1}$ at the 2$\sigma$ level \cite{Damour:1996zw}. 

As $\alpha$ affects the recombination history and diffusion damping of sound waves in the baryon-photon plasma, 
models with varying $\alpha$ can be probed using cosmic microwave background (CMB) anisotropies, now characterized at $\sim 0.1\%$ precision using data from the \textit{Planck} satellite \cite{Ade:2015xua,Aghanim:2018eyx} as well as a variety of ground-based experiments like the South Pole Telescope (SPT) \cite{Story:2012wx,Benson:2014qhw} and Atacama Cosmology Telescope (ACT) \cite{Louis:2016ahn}. These measurements require that the difference between the fine-structure constant today and at recombination obey the limit $\delta{\alpha}/\alpha\leq 7.3 \times 10^{-3}$ at 68\%~C.~L. \cite{Avelino:2001nr,Martins:2003pe,Rocha:2003gc,Ichikawa:2006nm,Menegoni:2009rg,2010PhRvD..82l3504G,2012PhRvD..85j7301M,Ade:2014zfo,deMartino:2016tbu,Hart:2017ndk}.

Measurements of the CMB anisotropies by \textit{Planck} provide an additional motivation for considering a spatially varying $\alpha$. Gravitational lensing smooths the CMB power spectrum while also inducing a non-Gaussian contribution to the trispectrum. It has been noted that the level of smoothing in the power spectrum is larger than what is expected given the measured amplitude of the non-Gaussian part of the trispectrum \cite{Ade:2015xua,Aghanim:2018eyx}. Analogous to the effects of weak gravitational lensing, the spatial variation of $\alpha$ smooths the CMB power spectrum and contributes to the non-Gaussian part of the trispectrum. 

It is thus possible that the measured smoothing/trispectrum discrepancy points towards modulation of the primordial CMB anisotropies beyond weak gravitational lensing at the level of about three standard deviations. In fact this possibility was extensively explored in Ref.~\cite{Aghanim:2018eyx}. There, the \textit{Planck} collaboration considered the effects of compensated isocurvature perturbations (CIPs) to explain this anomaly. A spatially varying $\alpha$ produces effects very similar to CIPs and may provide an alternative explanation. 

At a similar level of significance, the \textit{Planck} measurements  confirm a previously identified deviation from isotropy \cite{Ade:2015hxq}-  a `hemispherical asymmetry' in the CMB power spectra. The presence of a large-scale spatially varying $\alpha$, possibly correlated with the primordial anisotropies, may explain such an apparent deviation from isotropy in the CMB. 

Spatial fluctuations in $\alpha$  modulate the recombination history and rate of diffusion damping of baryon-photon plasma perturbations, and thus induce higher-order (and non-Gaussian) correlations in the CMB. This was pointed out in Ref.~\cite{Sigurdson:2003pd} and  applied to data in Ref.~\cite{OBryan:2013nip}, followed by Ref.~\cite{Ade:2014zfo}, in which the spatial dipole in $\alpha$ at recombination was directly limited to $C_{L=2}^\alpha \leq 1.3 \times 10^{-2}$ at 68\%~C.~L. For a scale invariant power spectrum this can be translated into a constraint to the amplitude of $A_{\rm SI}^\alpha = 2 C_{L=1}^{\alpha} < 2.6 \times 10^{-2}$. Here, we use 2015 \textit{Planck} satellite data (which includes small-scale polarization measurements) to test for spatial variation of the fine-structure constant on smaller scales ($\alpha$ multipoles $L\geq 8$). We find that the amplitude of a scale-invariant spectrum must have $A_{\rm SI}^\alpha < 8 \times 10^{-6}$ at 68\% C.~L. The dramatic improvement in the overall order-of-magnitude of the sensitivity results from the use of many more multipoles ($8\leq L\leq 100$) to search for $\alpha$ variations.
 
All attempts at using the CMB to search for the spatial variation of the fine-structure constant rely on a `separate-universe' (SU) approximation (for the response of CMB fluctuations to $\alpha$ variations) \cite{Sigurdson:2003pd,OBryan:2013nip,Ade:2014zfo}. Here we show that this approximation breaks down if the length scale of $\alpha$ fluctuations is smaller than the sound horizon at the surface of last scattering (SLS), analogous to an effect that occurs for compensated isocurvature perturbations \cite{He:2015msa}. This, in turn, implies that a more complete treatment of these types of effects may lead to additional sensitivity. 

This paper is organized as follows. In Sec.~\ref{sec:alphaCMB}, we explore the effect of varying $\alpha$ on the visibility function, and briefly describe how these changes propagate to the CMB power spectrum and then lay out the details of the relevant calculation using the codes \textsc{HyRec} \cite{HyRec} and \textsc{Camb} \cite{camb}.  We then determine how the additional non-Gaussian correlations induced by $\alpha$ fluctuations can be detected using an optimal estimator or existing CMB weak lensing data products.

In Sec.~\ref{sec:su_limits}, we use a toy model to show that the SU approximation should break down for $\alpha$-modulation on angular scales smaller than the acoustic horizon at the SLS ($L\gtrsim 100$). 

In Sec.~\ref{sec:current}, we compare this theory to data to search for spatial variation in $\alpha$ and generate constraints. At the current level of experimental precision, these (lensing-data derived) constraints are essentially optimal. In Sec.~\ref{sec:optimal}, we forecast the sensitivity of future experiments to $\alpha$ fluctuations using an optimal estimator. In Sec.~\ref{sec:discussion}, we explore the possibility that $\alpha$ fluctuations could explain apparent anomalies between theory and the observed amplitude of weak gravitational lensing the CMB. There, we also estimate the implications of our work for specific varying-$\alpha$ models, with an eye towards those that could explain the claimed dipole in $\alpha$ seen in observations of quasar spectra. We conclude in Sec.~\ref{sec:conclude}, summarizing our constraints and discussing how sensitivity to $\alpha$ fluctuations could be improved with novel cosmological observables.

We present some of the detailed expressions used in this paper in Appendix \ref{sec:details}. The second derivatives of CMB  power spectra, needed to construct $\alpha$-induced corrections  to the observed power spectra are described in Appendix \ref{sec:step_optimal}. 

\section{The CMB and the spatial variation of the fine-structure constant}
\label{sec:alphaCMB}

In this Section we summarize how the physics that produces the CMB depends on $\alpha$. After that we focus on how a spatially varying $\alpha$ affects the CMB. 

\subsection{An overview of how the CMB depends on the value of the fine-structure constant}

We briefly review how the spatial variation of the fine-structure constant affects the CMB. Readers interested in more details may find them in prior work, such as  Refs.~\cite{Sigurdson:2003pd,OBryan:2013nip,Hart:2017ndk}. The fine-structure constant sets the rates of processes relevant for hydrogen recombination \cite{HyRec} as well as the Thomson scattering cross section (which in turn sets the baryon-photon diffusion damping scale), and thus affects observable properties of the CMB \cite{1968ApJ...151..459S}.  

 \begin{figure*}
\begin{center}
\resizebox{!}{12cm}{\includegraphics{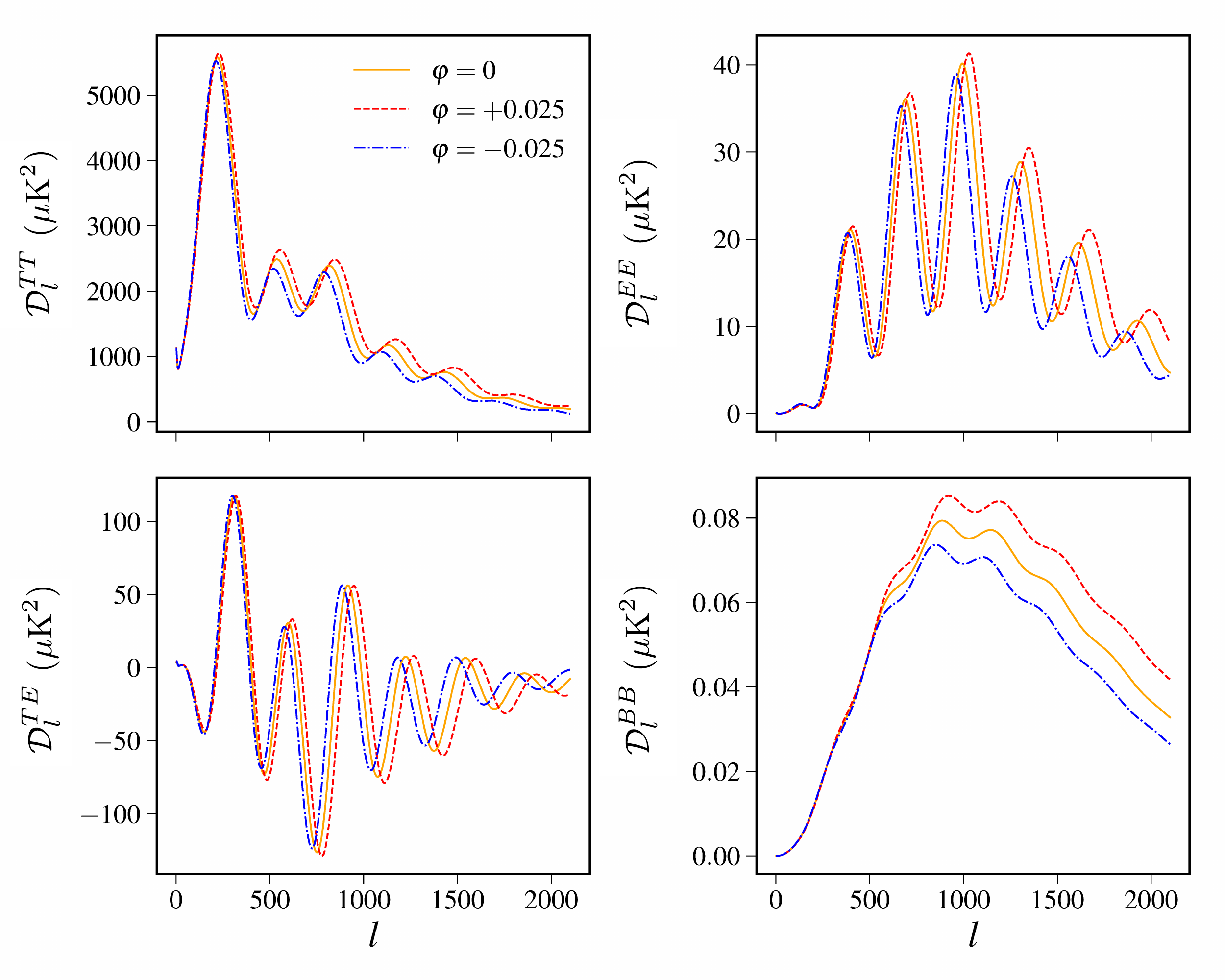}}
\caption{The change to the CMB power spectra as the fine-structure constant is varied, $\alpha = \alpha_0(1+\varphi)$. As discussed in the text an increased $\alpha$ leads to a shift in the peak of the visibility function to higher redshifts leading to an increase in the distance to the SLS which, in turn, shifts the angular scale of the CMB anisotropies to smaller scales (higher values of $l$). The earlier decoupling/recombination for higher $\alpha$ gives us a `snapshot' of the CMB anisotropies at a time when they are less damped, leading to an enhanced amplitude for scales above the damping scale. }
\label{fig:alphaCls}
\end{center}
\end{figure*} 

The impact of $\alpha$-modulation may be separated into effects induced by changes to the time-dependence of the decoupling process and effects induced by changes to perturbation evolution. This division is inspired by the line of sight integral used to compute the CMB anisotropies \cite{Seljak:1996is}. The line of sight integral allows us to write the anisotropy in temperature or polarization today in terms of a transfer function, $S_{\rm X}(k,\tau)$, as
\begin{equation}
X_{l}(k) = \int_0^{\eta_0} S_{\rm X}(k,\tau) j_l[k(\eta_0-\eta)] d\eta,
\end{equation}
where $\eta$ is the conformal time, $\tau$ is the optical depth, $X$ refer to temperature or E/B-mode polarization ($X\in {\rm T, E, B}$), respectively, $\eta_0$ is the conformal time today, and $j_l(x)$ is a spherical Bessel function of order $l$.  The measured power spectrum is then given by 
\begin{equation}
C_l = \frac{2}{\pi} \int k^2 dk P_\zeta(k) |X_{l}(k)|^2,
\end{equation}
where $P_\zeta(k)$ is the primordial power spectrum. 
The terms that appear in the source functions take the form 
\begin{equation}
 S_{\rm T,P}(k,\tau) \sim \mathcal{F}(\tau) \Delta_x(k,\eta; \tau).
 \end{equation}
 where $\mathcal{F}(\tau)$ is some function of the optical depth and $\Delta_x(k,\eta;\tau)$ is some linear perturbation either to a fluid component or to the gravitational potentials. Very roughly speaking we are separating the physics that determines what we `see' in the CMB from the physics that dictate the evolution of the photon perturbations of the CMB. A modulation of the fine-structure constant affects both of these aspects of the observed CMB. 

\begin{figure}
\begin{center}
\resizebox{!}{12cm}{\includegraphics{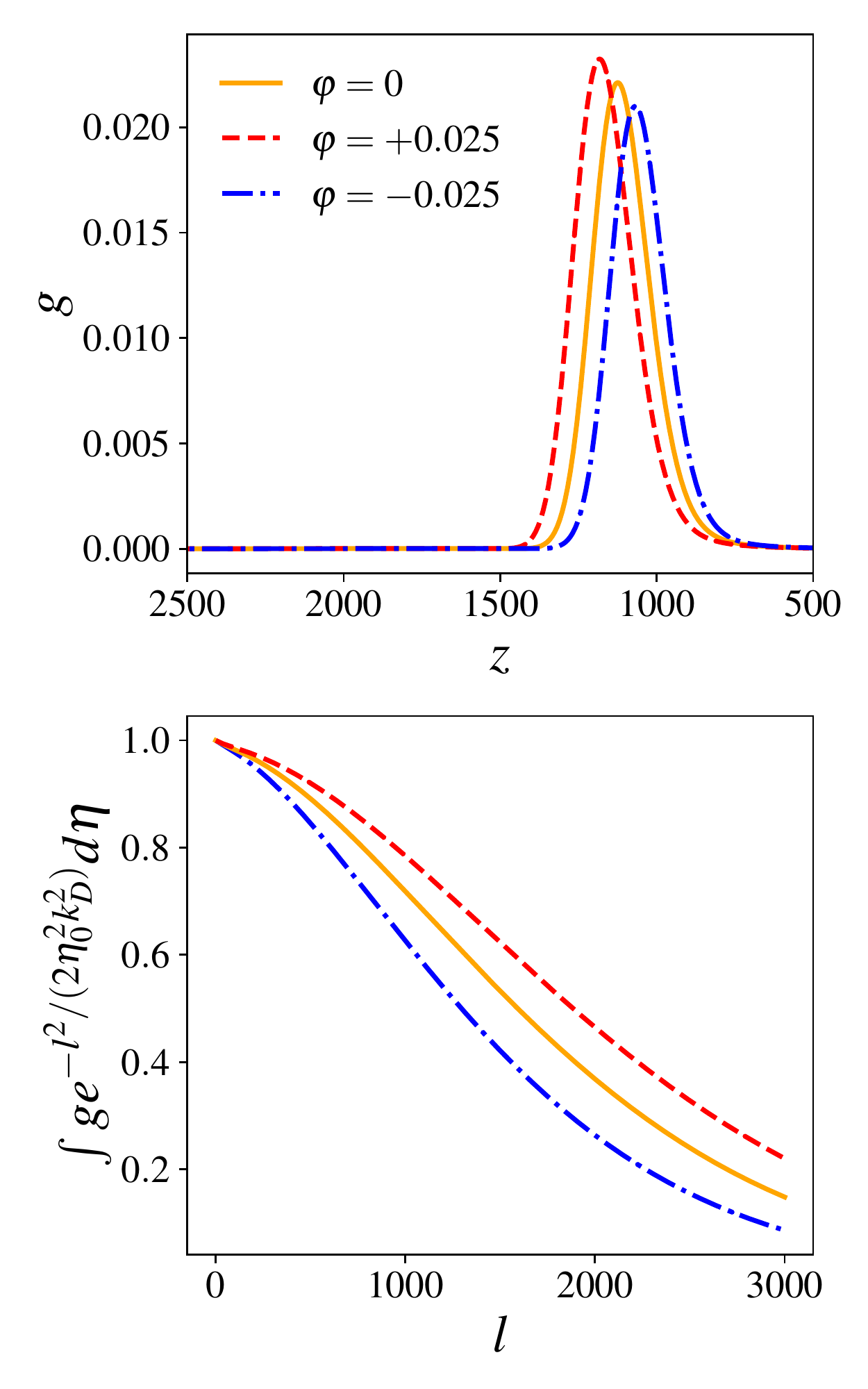}}
\caption{The change to the visibility function (top) and the damping of the CMB anisotropies (bottom) with a change in the fine-structure constant $\alpha = \alpha_0(1+\varphi)$. In the top panel we can see that an increase in $\alpha$ leads to stronger electromagnetic interactions shifting the peak of the visibility function (which marks the temperature at decoupling) to early times (i.e., higher redshift). An increase in $\alpha$ increases the Thomson scattering cross section leading to a larger differential cross section at early times. As we approach decoupling the additional changes to the ionization history causes a larger $\alpha$ to give a \emph{smaller} differential cross section. }
\label{fig:vismod}
\end{center}
\end{figure}

 A spatial modulation in the 
 fine-structure constant will cause a modulation to $\tau(\eta)$ through its effects on recombination \cite{Sigurdson:2003pd,Hart:2017ndk}.  As summarized in Ref.~\cite{Hart:2017ndk} several physical quantities that play essential roles in the physics of recombination depend on the fine-structure constant in a variety of ways:
\begin{align}
\label{eq:alpha_scaling}
\begin{split}
\sigma_{\rm T} \propto \alpha^2 
\qquad A_{2\gamma} &\propto \alpha^8 
\qquad P_{\rm S} A_{1\gamma} \propto \alpha^{6}
\\
~~\alpha_{\rm rec} \propto \alpha^2 
\qquad \beta_{\rm phot} &\propto \alpha^5 
\qquad T_{\rm eff} \propto \alpha^{-2},
\end{split}
\end{align} 
 where $\sigma_T$ is the Thomson scattering cross section; $A_{2 \gamma}$ is the two-photon decay rate of the second shell; $\alpha_{\rm rec}$ and $\beta_{\rm phot}$ are the effective recombination and photoionization rates, respectively; $T_{\rm eff}$ is the effective temperature at which $\alpha_{\rm rec}$ and $\beta_{\rm phot}$ are evaluated; and $P_S A_{1 \gamma}$ is the effective dipole transition rate for the main resonances. The overall effect of a modulation of $\alpha$ on recombination is to shift the peak and broaden the width of the visibility function, $g = \dot \tau e^{\tau}$ as shown in Fig.~\ref{fig:vismod}.
There we show the change to the visibility function when the fine-structure constant is shifted by a multiplicative factor, $\alpha = \alpha_0(1+\varphi)$. We can see that for values of $\alpha$ larger than the standard value the electromagnetic interactions are stronger leading to a shift in the peak of the visibility to earlier times.  In this sense the spatial modulation of $\alpha$ will cause the surface of last scattering to become `wrinkled'. 
 
The rate of change of the optical depth leads to a damping of anisotropies on scales below the diffusion scale.  This damping is controlled by the differential optical depth, $\dot \tau \equiv a n_H X_e \sigma_T$, where $n_e$ is the electron density and $X_e$ is the ionization fraction.  As we discuss further in Sec.~\ref{sec:su_limits}, the evolution of the temperature perturbations during the time when the differential optical depth is large compared to the Hubble rate, $ \dot \tau \gg \mathcal{H}$ (i.e., while baryons and photons are tightly coupled), gives a damped-driven harmonic oscillator equation of motion with a damping proportional to $1/\dot \tau$.  An increase in the fine-structure constant leads to a decrease in the damping and hence an increase in the overall amplitude of the power spectra. We show the change in the level of diffusion damping for different values of $\alpha$ in the bottom panel of Fig.~\ref{fig:vismod}. 

 \begin{figure*}
\begin{center}
\resizebox{!}{9.2cm}{\includegraphics{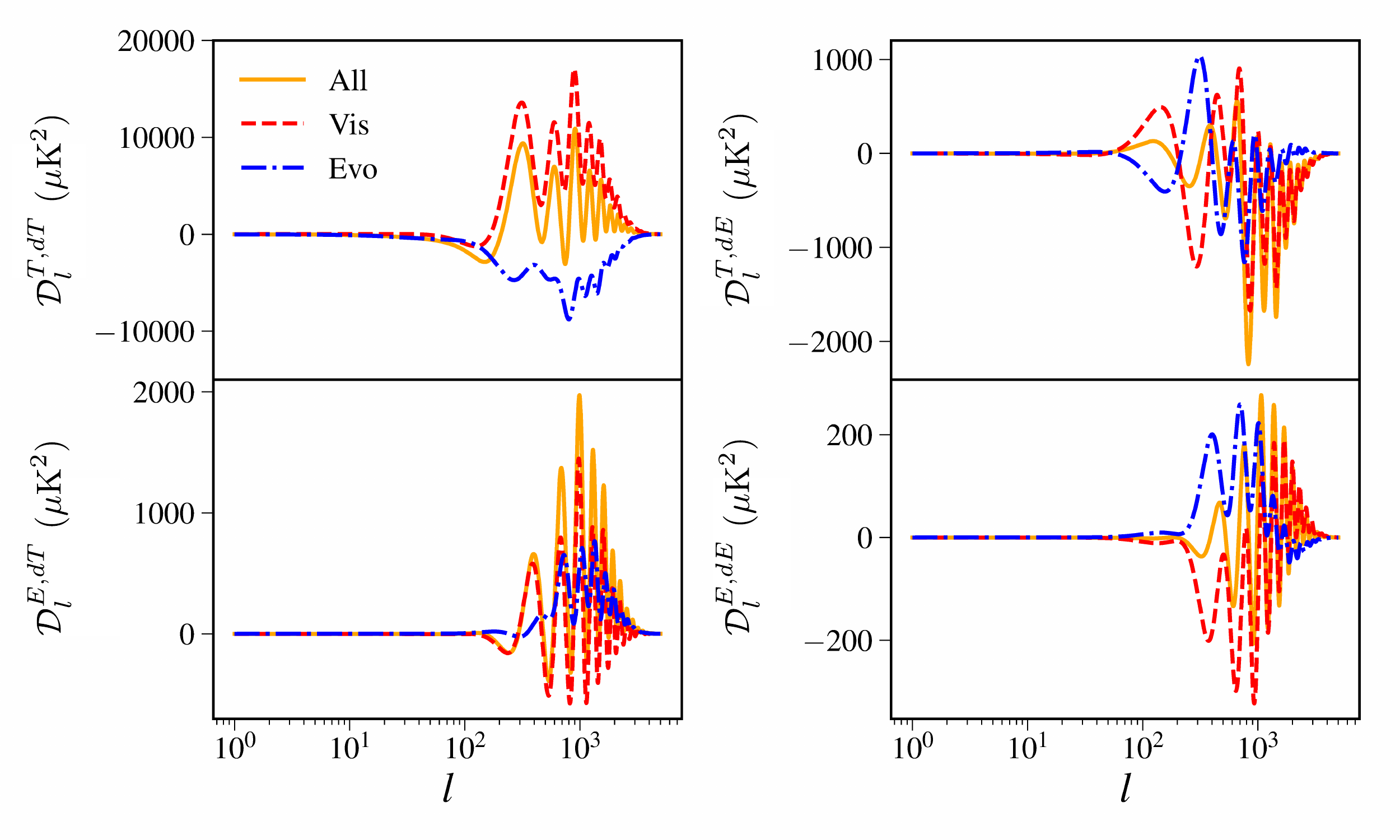}}
\caption{The relative contribution of variations in the visibility function (`Vis')  and in the transfer function (`Evo') to the overall derivative power spectra. It is clear that $\alpha$-modulation has a significant impact on both terms.}
\label{fig:comp1}
\end{center}
\end{figure*} 

The $\alpha$-modulation of the distance to the SLS and the diffusion damping contribute nearly equally to the full modulation.  We can see this in Fig.~\ref{fig:comp1}, which shows their relative contribution to the cross-power spectrum between the temperature or polarization and the derivative of the temperature or polarization with respect to $\varphi$: $\mathcal{D}_l^{{\rm X,dY}} \equiv l(l+1) C_l^{{\rm X,dY}}/(2\pi)$. The effect of $\alpha$-modulation on the evolution of the perturbations we see in the CMB has an important impact on the way in which we construct estimators for the $\alpha$ modulation as we discuss in detail in Secs.~\ref{sec:tri} and \ref{sec:su_limits}.

\subsection{Characterizing spatial fluctuations in the fine-structure constant}

We will parameterize the fluctuations in the fine-structure constant as 
\begin{equation}
\alpha(\vec x) = \alpha_0[1+\varphi(\vec x)].
\end{equation}
In theories which promote the fine-structure constant to a dynamical quantity, the field $\varphi$ will also depend on time. Previous work has used the CMB to constrain this time-evolution and has found that the fine-structure constant cannot vary appreciably during the process of recombination (In Ref.~\cite{Hart:2017ndk}, it is shown that $\Delta \eta_* |\dot{\alpha}/\alpha| \lesssim 10^{-3}$ where $\Delta \eta_*$ is the duration of decoupling- i.e.~the width of the visibility function). Therefore we take $\dot \alpha =0$ here and leave simultaneous constraints to both temporal and spatial variation of $\alpha$ to future work. 

We write the Fourier transform of $\varphi$ as 
\begin{equation}
\varphi(\vec x) = \int \frac{d^3 k}{(2\pi)^{3}} \varphi(\vec k) e^{i \vec k \cdot \vec x},
\end{equation}
and define its power spectrum in the usual way, 
\begin{equation}
\langle \varphi(\vec k) \varphi^*(\vec k\ ')\rangle = (2\pi)^3 \delta_D^{(3)}(\vec k - \vec k\ ')P_{\alpha}(k).
\end{equation}
In the limit of instantaneous decoupling, the $\alpha$-modulation angular distribution at the SLS is given by 
\begin{equation}
\varphi_{L M} =  \frac{4 \pi i^L}{(2\pi)^{3/2}}\int d^3 k \varphi(\vec k) j_L(k \chi_*) Y^*_{L M}(\hat k),
\end{equation}
which gives rise to an angular power spectrum at the SLS (which is located at a comoving distance $\chi_*$) of
\begin{eqnarray}
C_L^{\alpha} &=& \langle \varphi_{L M} \varphi^*_{L M}\rangle\\ &=&   \frac{A_\alpha  \Gamma \left(L+\frac{n_\alpha}{2}\right)}{\Gamma \left(L-\frac{n_\alpha}{2}+2\right)},\\
&\simeq& \frac{A_\alpha}{[L (L+1)]^{1-n_\alpha/2}},\ {\rm for}\ L\gtrsim 1,
 \label{eq:ang_delta_sls_ps}
\end{eqnarray}
where $A_\alpha \equiv \mathcal{A}_\alpha \Gamma \left(1-\frac{n_\alpha}{2}\right)/\Gamma \left(\frac{3}{2}-\frac{n_\alpha}{2}\right)$ and where 
we assume that the $\alpha$-modulation power spectrum is a power law: $P_{\alpha}(k) = \mathcal{A}_\alpha(k \chi_*)^{n_\alpha}/k^3 $. Note that for a scale invariant spectrum we have $C_L^{\alpha} = A^{\rm SI}_\alpha/[ L(L+1)]$; a white-noise angular power spectrum with $C_L^{\alpha} =A^{\rm WN}_\alpha$ corresponds to the limit $n_\alpha\rightarrow 2$. If $\varphi$ couples to the inflaton field or standard model particles (as it does in most theories- i.e., Ref.~\cite{Barrow:2013uza}) then we expect that its power spectrum is scale invariant. 

Using the power spectrum we can compute the relationship between the variance of $\alpha$ on the sky and the amplitude of its power spectrum:
\begin{eqnarray}
\frac{\sigma_{\alpha}^2}{\alpha_0^2} &=& \frac{1}{4\pi} \int \langle \varphi(\hat n)^2 \rangle d^2 \Omega = \sum_{L=1}^{L_{\rm max}} \frac{2L+1}{4\pi} C_L^{\alpha}.\label{eq:normps}
\end{eqnarray}
As has been pointed out, when calculating the effects of $\varphi \neq 0$ on the CMB we truncate the sum over the $\varphi$ multipoles at $L_{\rm max}=100$ since the modulation of CMB fluctuations is suppressed for $\varphi$ modes with wavelengths smaller than the acoustic horizon at the SLS (for details see Sec.~\ref{sec:su_limits}). 

The spatial fluctuations in the fine-structure constant could have different statistical properties, such as a Gaussian power spectrum \cite{Sigurdson:2003pd}, or a white-noise (constant) power spectrum \cite{OBryan:2013nip}. In this work, we focus on the scale-invariant case, but also place constraints on the white-noise case to allow a comparison with Ref.~\cite{OBryan:2013nip}.

\subsection{The effects of a spatially-varying fine-structure constant on the CMB power spectrum}
\label{sec:eff}

As discussed in Sec.~\ref{sec:alphaCMB} the spatial modulation of the fine-structure constant affects both the location of the SLS as well as the evolution of the perturbations. A full accounting for the effects of $\varphi(\vec x)$ requires a solution to the modified set of second-order evolution equations. As a first approximation to this, we use the SU approximation and take the solution to the original evolution equations (where $\alpha$ does not vary in space) and then expand that solution in a power-series in $\varphi(\vec x)$. The SU approximation has been used to calculate the effects of compensated isocurvature perturbations (CIPs) in which the initial baryon number density fluctuates in space with an equal and opposite CDM fluctuation \cite{Grin:2011nk,Grin:2011tf, Grin:2013uya,He:2015msa,Smith:2017ndr}. 

The estimators that we construct using the SU approximation are limited in that they effectively filter the data and remove information about the modulation on small scales. The exact cut-off in this filter is determined by a comparison between the power-series approximation and a perturbative solution to the full dynamical equations in the presence of the modulation. For example, CIPs effectively cause a modulation in the baryon-photon sound speed, causing a spatial modulation in the sound-horizon of the CMB. A careful analysis using a simplified model for the evolution equations describing the temperature perturbations in Ref.~\cite{He:2015msa} showed that the resulting estimator filtered out information on the CIPs on scales smaller than the sound horizon of the CMB. We perform a similar analysis for the spatial modulation of $\alpha$ in Sec.~\ref{sec:su_limits} and find that the power law expansion is accurate for $L<L_{\rm max} \simeq 100$. Unless explicitly noted otherwise, we use a maximum value of $L_{\rm max}=100$ everywhere.

We now present results for CMB temperature and polarization anisotropies in the presence of $\alpha$ fluctuations. The formalism of this section is nearly identical to that employed in Ref.~\cite{Smith:2017ndr} in which we used measurements of the lensing potential power spectrum to place constraints on compensated isocurvature perturbations. Here we will leave out the details and direct interested readers to Ref.~\cite{Smith:2017ndr}.

Weak gravitational lensing and the spatial modulation of the fine-structure constant can be thought of as a modulation of a `background' CMB anisotropy $T(\hat n)$ yielding an observed anisotropy $T_{\rm obs}(\hat n)$.\footnote{Although here we focus on the temperature anisotropies, the results we present also apply to measurements of the polarization of the CMB.}  
In the presence of both weak gravitational lensing, with lensing potential $\phi(\hat{n})$, and fractional $\alpha$-variation $\varphi(\hat{n})$, the temperature anisotropies are approximately given by 
\begin{align}
T_{\rm obs}(\hat n) =&T\left[\hat n + \vec{\nabla} \phi(\hat n), \varphi(\hat n)\right] \\
\simeq&  T(\hat n) + \nabla_i \phi \nabla^i  T + \varphi(\hat n) \frac{\partial  T}{\partial \varphi}(\hat n)\bigg|_{\varphi = 0}\nonumber \\ +& \frac{1}{2}\left( \nabla_i \phi \nabla_j \phi \nabla^i \nabla^j  T + \varphi^{2}(\hat n) \frac{\partial^2  T}{\partial \varphi^2}(\hat n)\bigg|_{\varphi = 0}\right) + \cdots, \label{eq:rspaceexp}
\end{align} 
where the terms proportional to derivatives of $\phi(\hat{n})$ are standard lensing contributions \cite{Zaldarriaga:1998te,Hu:2000ee}. Also note that we neglect any cross terms of the sort $\varphi \nabla \phi$ since we assume that $\varphi$ does not have any correlation with other cosmological fields. As such, the upper limits presented here are conservative relative to models which predict a correlation between $\varphi$ and other fields. 
Additionally, one must include a noise term, so that the total observed temperature at each point on the sky can be written $T^t(\hat n) = T_{\rm obs}(\hat n) + T^N(\hat n)$, where we assume that we are using beam-deconvolved maps. This leads to an estimated power spectrum for the beam-deconvolved map \cite{Knox:1996cd}
\begin{equation}
C_{l}^{{\rm TT},{\rm t}} = C_{l}^{{\rm TT},{\rm obs}} + C_{l}^{{\rm TT},{\rm N}}.
\label{eq:obsmaps}
\end{equation}

From this  it is straightforward to show that in the presence of both lensing and a spatially varying $\alpha$ the observed power spectrum becomes
\begin{equation}
C^{\rm TT, {\rm obs}}_l = \tilde{C}^{\rm TT}_l+\delta C^{\rm TT,\phi}_l + \delta C^{\rm TT,\alpha}_l
\label{eq:Clexpand},
\end{equation}where $\tilde{C}^{\rm XX'}_l$ denotes the true primordial ${\rm X}{\rm X}^{\prime}$ power spectrum (without corrections from noise, $\varphi$ fluctuations, or gravitational lensing) of the quantity ${\rm X}$, which can denote temperature or E/B-mode polarization moments ($X\in\{{\rm T,E,B}\}$ ). The standard lensing correction to the power spectrum is denoted by $\delta C^{\rm TT,\phi}_l$, and is computed using the usual techniques from Ref.~\cite{Okamoto:2003zw}. 

The correction to the TT power spectrum from $\alpha$ fluctuations ($\delta C^{\rm TT,\alpha}_l$) is computed using a formalism first developed for the flat-sky approximation in Ref.~\cite{Sigurdson:2003pd} and then generalized to the whole sky in Ref.~\cite{OBryan:2013nip}. Here we adopt the notation of Refs.~\cite{He:2015msa,Smith:2017ndr}, developed for CIPs, but replace baryon-density derivatives with derivatives with respect to $\varphi$. It is then straightforward to apply Eq.~(\ref{eq:rspaceexp}) to obtain
\begin{equation}
\delta C^{\rm TT,\alpha}_l\simeq \sum_{Ll'} C_L^{\alpha} C^{\rm dT,dT}_{l'} (K_{ll',0}^L)^2 G_{Ll'} + \frac{\sigma_{\alpha}^2}{\alpha_0^2} C_l^{\rm T,d^2T}, \label{eq:grinTTobs}
\end{equation}
where
\begin{eqnarray}
C_l^{\rm dX,dX} &\equiv& \frac{2}{\pi} \int k^2 dk P_{\zeta}(k) \left(\frac{dX_l(k)}{d\varphi}\right)^2,\label{eq:obs_tta}\\
C_l^{\rm X,d^2X} &\equiv& \frac{2}{\pi} \int k^2 dk P_{\zeta}(k) X_l(k) \frac{d^2X_l(k)}{d\varphi^2},\label{eq:obs_ttb}\\
G_{L,l'} &\equiv& \frac{(2L+1)(2l'+1)}{4\pi}, \\
K_{ll',s}^L &\equiv& \left(\begin{array}{ccc}l & L & l' \\s & 0 & -s\end{array}\right),
\end{eqnarray}where $P_{\zeta}(k)$ is the usual power spectrum of primordial curvature fluctuations and $X_{l}(k)$ is the usual CMB transfer function mapping $\zeta$ fluctuations in $k$-space to angular fluctuations in CMB observables.

Since the effects of a spatially-varying $\alpha$ on CMB anisotropies occur mainly at the SLS it follows that $C^{\rm dT,dT}_{l}$ is only significant on scales smaller than the acoustic horizon, $l \gtrsim 100$ (see also Fig.~\ref{fig:Clderivs}). Furthermore, for a scale-invariant power spectrum peaks at small $L$, leading to a separation of scales which allows us to write the effects of $\alpha$-variation on the observed CMB power spectrum as \cite{Munoz:2015fdv,Smith:2017ndr}
\begin{eqnarray}
\delta C^{\rm XX',\alpha}_l &\simeq& \frac{1}{2} \frac{\sigma_{\alpha}^2}{\alpha_0^2} \frac{\partial^2 C_l^{\rm XX'}}{\partial \varphi^2}\bigg|_{\varphi = 0},\label{eq:flatClobs}
\end{eqnarray}
where $X$ and $X'$ can be T or E. 

In the absence of primordial gravitational waves the $\alpha$ variation transforms E-mode polarization into B-mode polarization (in a process that is analogous to the generation of B-mode polarization through lensing).  In this case the induced B-mode polarization is  \cite{Grin:2011tf}
\begin{equation}
C_l^{\rm BB} \simeq C_l^{\rm BB,\ {\rm lensed}}+\sum_{L,l'}^{L+l'+l~{\rm odd}}C_L^{\alpha}  C_{l'}^{\rm dE,dE} (K^L_{ll',2})^2\label{eq:bbind}.
\end{equation}
The expression for the induced B-mode polarization cannot be written in the form of Eq.~(\ref{eq:flatClobs}: The second term in Eq.~(\ref{eq:grinTTobs}) arises because there is a non-zero value for the temperature anisotropy even in the absence of CIPs (and analogously so for the E-mode polarization anisotropy). On the other hand, there is no B-mode in the unlensed CMB in the absence of primordial gravitational waves, and thus no term like the second term of Eq.~(\ref{eq:grinTTobs}) in the expression for the CIP-induced B-mode power spectrum. The absence of this term prevents the algebraic simplifications that yield Eq.~(\ref{eq:flatClobs}.

\begin{figure*}
\begin{center}
\resizebox{!}{12cm}{\includegraphics{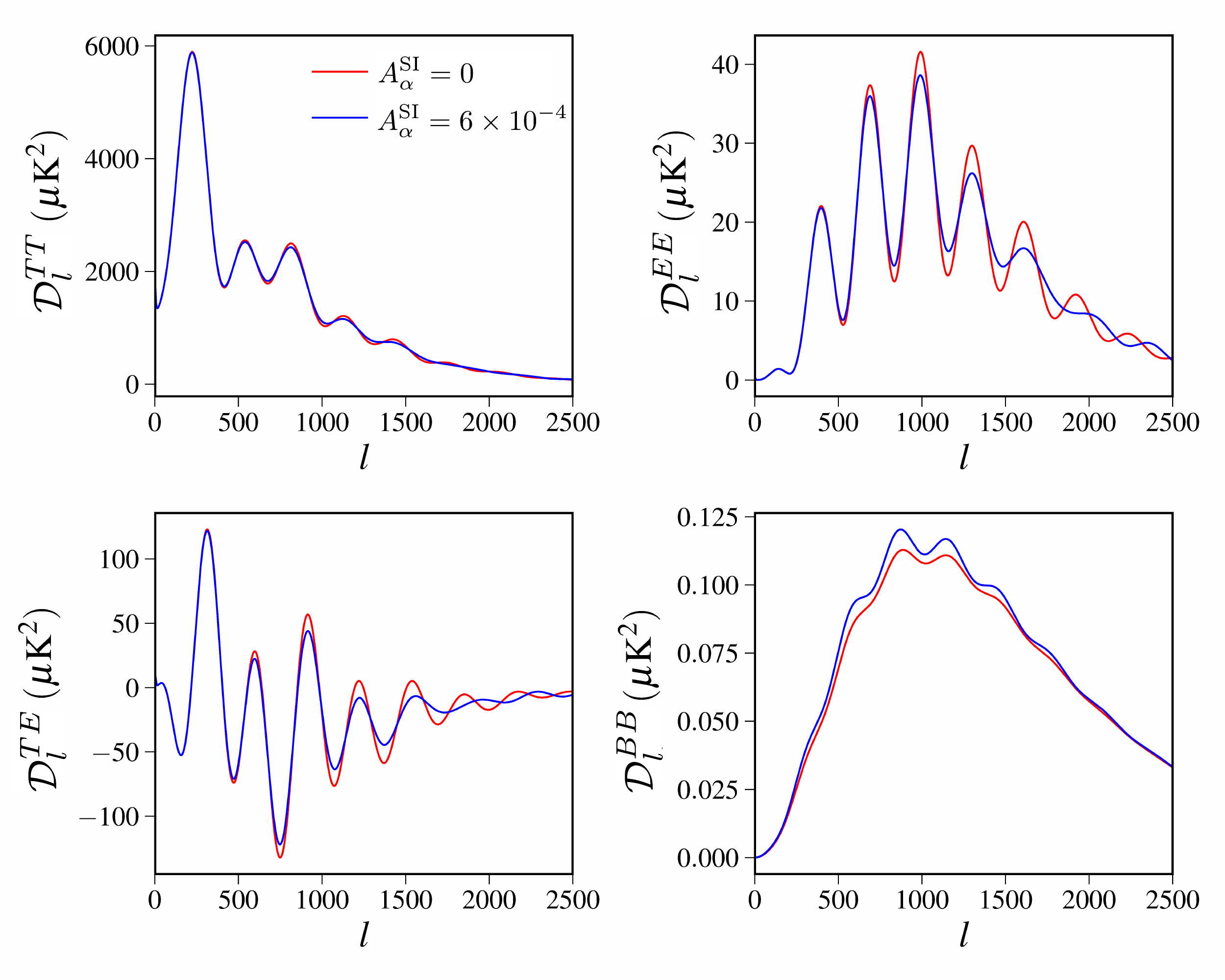}}
\caption{The change to the CMB power spectra due to stochastic scale-invariant fluctuations in the fine-structure constant (we have set the amplitude, $A_\alpha^{\rm SI}$, to be a large enough to see the effects). The spatial variation of $\alpha$ causes a smoothing of the anisotropies and affects the polarization more than it does the temperature power spectrum. Note that the B-mode power spectrum does not include the effects of gravitational waves but does show the B-modes generated by lensing of the E-mode polarization and spatial fluctuations in $\alpha$.}
\label{fig:alphaClsRMS}
\end{center}
\end{figure*}

We show the unmodulated and modulated power spectra in Fig.~\ref{fig:alphaClsRMS}. There we can see that a spatially varying fine-structure constant leads to a smoothing of the CMB power spectra and has a larger effect on the polarization than it does on the temperature anisotropies. Also note that we have chosen $A^{\rm SI}_\alpha = 6 \times 10^{-4}$ for this figure in order to highlight the power spectrum modulation; when saturating the upper limit using the T and E spectra we have $A_\alpha^{\rm SI} <5.2 \times 10^{-5}$  (see Sec.~\ref{sec:current})

Since the power spectrum constraints are driven by a modulation of the temperature and E-mode polarization Eqns.~(\ref{eq:Clexpand}) and (\ref{eq:flatClobs}) give us an efficient way to compute the effects of a spatially modulated fine-structure constant on the CMB power spectra. When computing the modulated power spectrum for the Markov chain Monte Carlo (MCMC), we use a finite difference to compute the second derivative of the power spectra with respect to $\varphi$ with a stepsize of $\Delta \varphi = 0.012$. We show that this step size gives the best estimate of the second derivative in Appendix \ref{sec:step_optimal}. 

Computing these derivatives requires the evaluation of CMB power spectra $C_{l}^{\rm XX'}$ for various global $\varphi$ values. To compute these, we modify the \textsc{Camb} \cite{camb} code. We set \textsc{Camb} to use the recombination code \textsc{HyRec} \cite{AliHaimoud:2010ab,HyRec}, in order to include completely the rich ensemble of effects relevant for adequately modeling cosmic recombination in the precision cosmology era. \textsc{HyRec} has the added advantage that $\varphi$ may be readily changed by changing a single argument, ensuring that all the relevant derivatives are correctly computed without neglecting any relevant physical effects. 

\subsection{Additional effects of a spatially-varying fine-structure constant on the CMB}
\label{sec:tri}
The spatial modulation of the fine-structure constant also produces a contribution to correlations beyond the CMB power spectrum. In particular, a non-zero realization of $\varphi(\vec{x})$ induces off-diagonal correlations between CMB multipole moments with $l\neq l^{\prime}$ and $m\neq m^{\prime}$ \cite{Sigurdson:2003pd,OBryan:2013nip}. Using these, an optimal estimator for $\varphi_{LM}$ may be constructed, which in turn can be used to construct an optimal estimator of $A_{\alpha}$. The formalism is very similar to that used in weak lensing of the CMB \cite{Okamoto:2003zw}. We review these correlations in Sec.~\ref{sec:offd}, discuss the related optimal estimators in Sec.~\ref{sec:est} and then in Sec.~\ref{sec:lenscon} obtain a method for applying existing CMB lensing measurements to probe spatial variations in $\alpha$. 

\subsubsection{Off-diagonal correlations}
 \label{sec:offd}
Deflections of CMB photons and higher-order modulations of the transfer functions produce off-diagonal CMB correlations for fixed lens and $\varphi$ realizations, given by \cite{Okamoto:2003zw,Grin:2011nk,Grin:2011tf} 
\begin{eqnarray}
\VEV{X_{l m} X'_{l'm'}}&\big|_{\rm lens,\alpha}& = \tilde{C}_l^{X X'} \delta_{l l'} \delta_{m -m'} (-1)^m\nonumber \\ &+& 
\sum_{LM} (-1)^M \wigner l m {l'} {m'} L {-M} \nonumber \\ &\times&\bigg[\phi_{LM} f^{XX'}_{l Ll'}+ \varphi_{LM} h^{XX'}_{l L l'}\bigg],
\label{eq:2pt}
\end{eqnarray}
 where $f^{XX'}_{l L l'}$ and $h^{XX'}_{l L l'}$  are the lensing/$\alpha$ response functions for different quadratic pairs (see Table \ref{tab:CIP_resp} in Appendix \ref{sec:details}) and are defined in terms of the unmodulated power spectrum, $\tilde{C}_l^{XX'}$, the appropriately weighted Wigner coefficients and the derivative power spectra \begin{equation}
C_l^{\rm X,dX'} \equiv \frac{2}{\pi} \int k^2 dk P_{\zeta}(k) X_{l}(k)\frac{dX'_l(k)}{d\varphi}.\label{eq:estnumdir} \end{equation}
The multipole moments of the lensing-potential are denoted by $\phi_{LM}$. This formalism was first developed for $\alpha$ fluctuations in Ref.~\cite{OBryan:2013nip}. Here we use the equivalent notation of Ref.~\cite{Smith:2017ndr}.

\subsubsection{Minimum-variance estimators for $\phi_{LM}$ and $\varphi_{LM}$.}
\label{sec:est}

Under the null hypothesis (i.e., no $\alpha$ variation), the minimum-variance estimator for the `deflection field'
$d_{LM}^{\omega}\equiv\sqrt{L(L+1)}\phi_{LM}$ from a single pair $\omega=XX'$ of observables is \cite{Okamoto:2003zw,Namikawa:2011cs},
\be
\hat d^{\omega}_{LM} = A_L^\omega \sum_{l m, l'm'} \left(-1\right)^{M} X_{l m} X'_{l' m'} \wigner l m {l'} {m'} L {-M} g^{\omega}_{l l'L},
\label{eq:dhat}
\ee
where $A_L^\omega$ is the normalization and $g^\omega_{l l'L}$ is the optimal weights and are defined in Appendix \ref{sec:details}.
This in turn can be used to derive (in the absence of a non-zero $\varphi$) an optimal estimator for the lensing-potential power spectrum
\be
\hat C_L^{\phi \phi} = \frac{1}{2L+1} \sum_{\omega,\beta}\sum_{M=-L}^L v^\omega_{L}v^\beta_{L}  \frac{\hat{d}^\omega_{LM}\hat{d}^{*\beta}_{LM}}{L(L+1)}-B_L,
\label{eq:CIPestp}
\ee
where $v^\omega_{L}$ 
are weights chosen to yield an optimal estimator for the deflection field and $B_L$ are the standard Gaussian noise bias and non-Gaussian lensing bias contributions to the CMB four-point correlation \cite{PhysRevD.67.123507,2011PhRvL.107b1301D,Ade:2015zua}. See Appendix \ref{sec:details} for detailed formulae. 

\textit{Mutatis mutandis} the off-diagonal correlations induced by $\varphi$ in Sec.~\ref{sec:offd} may be used to obtain a minimum-variance estimator of $\varphi_{LM}$ (as derived for CIPs in Refs.~\cite{Grin:2011tf,He:2015msa} and in different notation for $\alpha$ fluctuations in Ref.~\cite{OBryan:2013nip}). The derivation in Ref.~\cite{He:2015msa} closely follows the treatment in Ref.~\cite{Namikawa:2011cs} and generalizes to $\alpha$ fluctuations.

\subsubsection{Contribution of fine-structure constant fluctuations to CMB lensing estimators}
\label{sec:lenscon}

As discussed, estimates of the lensing-potential power spectrum, $C_L^{\phi \phi}$, are built out of the (non-Gaussian) connected part of the CMB trispectrum \cite{Okamoto:2003zw,2011PhRvL.107b1301D,Ade:2015zua}.  In the presence of a spatially varying $\alpha$, the estimator used to reconstruct the lensing-potential power spectrum gains an additional contribution proportional to $A_\alpha$, a bias which can itself be used to estimate $A_{\alpha}$ using existing lensing data products. The method presented here closely follows the method used in Ref.~\cite{Smith:2017ndr} with the CIP modulation field, $\Delta$, replaced by the $\alpha$ modulation field $\varphi$. We will summarize the important points here and direct the reader to Ref.~\cite{Smith:2017ndr} for details.

With a non-zero spatial modulation of $\alpha$ it is straightforward to show that the standard lensing estimator gains a contribution from $C_L^{\alpha}$:
\begin{eqnarray}
\VEV{\hat C_L^{\phi \phi}} 
&=&  C_L^{\phi \phi} + C_L^{\alpha} \mathcal{F}_L,
\label{eq:Alphacontrib}
\end{eqnarray}
where
\begin{eqnarray}
\mathcal{F}_L&\equiv&\sum_{\omega,\beta} w_{\omega} w_{\beta}Q^\omega_LQ^\beta_L,\\
Q^\omega_L &\equiv& \dfrac{\sum_{l l'} h^\omega_{l L l'} g^\omega_{l L l'} }
{\sum_{l l'} f^\omega_{l L l'} g^\omega_{l L l'}},
\end{eqnarray}
and as before, $h^\omega_{l L l'}$, $f^\omega_{l L l'}$ are listed in Table \ref{tab:CIP_resp} in Appendix \ref{sec:details}.  The brackets in Eq.~(\ref{eq:Alphacontrib}) denote an average over realizations of the primordial CMB, realizations of the lensing potential $\phi$, and realizations of the fine-structure modulation $\varphi$.

\begin{table}[hbtp!]
		\begin{tabular}{ l  c  c  }
			\hline
			\hline
			$L$ &  \textit{Planck} & CMB-S4   \\             
			\hline
			1 & 2.83 & 0.384   \\
			2 & 2.3 & 1.06  \\
			3 & 2.49 & 1.23  \\
			4 & 2.67 & 1.35  \\
			5 & 2.81 & 1.44  \\
			6 & 2.93 & 1.51 \\
			7 & 3.02 & 1.56  \\
			8 & 3.09 & 1.61  \\
			9 & 3.15 & 1.64  \\
			10 & 3.2 & 1.68  \\
			$>$10 & 3.29 & $1.21 L^{0.143}$  \\
			\hline
		\end{tabular}
		\caption{The $\alpha$-modulation contribution to the lensing potential power spectrum estimator, $L^4\mathcal{F}_L$, defined in Eq.~(\ref{eq:Alphacontrib}). The $\alpha$-modulation contribution is different for the two instruments because its affect on the lensing potential power spectrum estimator depends on the noise properties of the instrument.}
		\label{tab:CIP}
	\end{table}

 \begin{figure}[ht]
\begin{center}
\resizebox{!}{9.25cm}{\includegraphics{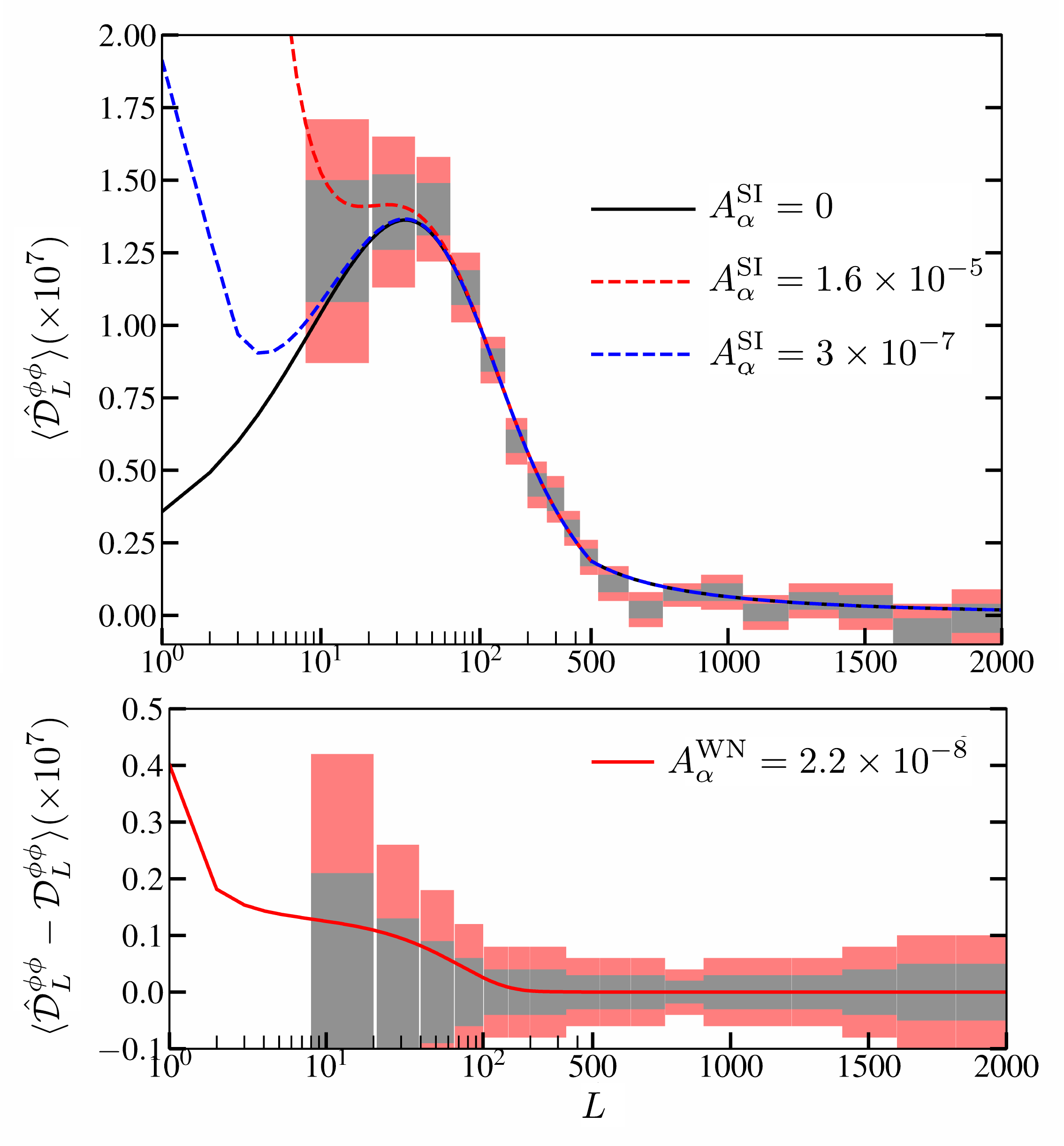}}
\caption{The affects of $\alpha$ modulation on the lensing potential power spectrum estimator. The top panel shows how a scale-invariant $\alpha$ modulation contribution. The lower panel shows the residual lensing potential power spectrum estimator in the presence of a white-noise $\alpha$ power spectrum. The amplitudes have been chosen to saturate the 95\% C.L.~bounds discussed in Sec.~\ref{sec:current}.}
\label{fig:Clpp}
\end{center}
\end{figure} 

The optimal weights in Eq.~(\ref{eq:oweight}) show that the $\varphi$ contribution to the lensing potential power spectrum depends on the noise properties of the relevant CMB experiment.  The full \textit{Planck} lensing analysis \cite{Ade:2015zua}, which includes both CMB temperature and polarization maps, computes a minimum variance (mv) estimator from all possible CMB map auto- and cross-correlations, as discussed in detail in Sec.~\ref{sec:est}. The \textit{Planck} lensing estimator relies on maps constructed from the 143 GHz and 217 GHz channels. 
\begin{table}[!ht]
\begin{center}
\noindent\adjustbox{max width=1.2\textwidth}{%
\begin{tabular}{p{15mm}ccc}
\hline\hline
Channel & $\theta$ (arcmin) & $w_T\ (\mu{\rm K}\ {\rm arcmin})$ & $w_P\ (\mu{\rm K}\ {\rm arcmin})$ \\\noalign{\smallskip}\hline\noalign{\smallskip}
143 GHz & 7 & 30 & 60  \\\noalign{\smallskip}
217 GHz & 5 & 40 & 95\\\noalign{\smallskip}
CMB-S4 & 3 & 1 & 1.4
\\\noalign{\smallskip}
\hline\hline
\end{tabular}}
\caption{Planck sensitivity in the 143 and 217 GHz channels to temperature and polarization at the two frequencies used to estimate the lensing-potential \cite{Ade:2015zua,Adam:2015vua}. The last line gives the sensitivity for CMB-S4, a proposed next generation CMB telescope \cite{Abazajian:2016yjj}. \label{table:plancksens}} 
\end{center}
\end{table}
The \textit{Planck} analysis also uses a bandpass filter in harmonic space to restrict the power spectrum multipoles to $100 \leqslant  l \leqslant 2048 $. We also list the noise parameters associated with the fourth-generation CMB experiment, CMB-S4 \cite{Abazajian:2016yjj}. We compute the sensitivity of CMB-S4 to a spatially varying fine-structure constant in Sec.~\ref{sec:optimal}.  We show the $\alpha$-modulation contribution to the lensing potential power spectrum estimator, $\hat{\mathcal{D}}_l^{\phi \phi} \equiv [l(l+1)]^2C_l^{\phi \phi}/(2\pi)$ in Fig.~\ref{fig:Clpp}. 

\section{Assessing the separate universe approximation}
\label{sec:su_limits}
We have investigated the effects of spatial variation of the fine-structure constant by using the SU approximation, where we calculate the effects of the spatial variation of the fine-structure constant on the CMB by computing a power series of the CMB maps in $\varphi$. 

This power series would be accurate to all scales if the modulation due to $\varphi(\vec x)$ only affected the visibility function, which codifies the projection of dynamical perturbations onto a fixed-redshift `screen,' but as we now discuss, the dynamical equations are also affected. In the case of compensated isocurvature perturbations in Ref.~\cite{He:2015msa}, such complications imply that the power law expansion is only accurate down to some cut-off scale $L_{\rm max}$. Here we determine this cut-off for spatial modulation of the fine-structure constant. 

To a first approximation the observed CMB provides a `snapshot' of the temperature and polarization perturbations around the SLS:
\begin{eqnarray}
T(\hat n) &=& \int_0^{\eta_0}  T(\chi, \chi \hat n, \eta)  g(\eta) d \eta,\\
E(\hat n) &=&- \int_0^{\eta_0} g(\eta) \frac{\dot{T}(\chi,\chi\hat n,\eta)}{\dot{\tau}(\eta)}d\eta,
\end{eqnarray}
where $\chi = \eta_0 - \eta$ is the comoving distance, $g(\eta)$ is the visibility function, T are the temperature perturbations, and E are the E-mode polarization perturbations.

We will consider the evolution of the temperature and polarization perturbations using a tight-coupling approximation, where $\dot \tau \gg \dot a/a$. In this case the temperature perturbations follow an equation of motion (neglecting super-horizon terms, polarization source terms, and other complications)
  \begin{eqnarray}
  \ddot{T} - c_s^2 \nabla^2 T &-&2\beta \nabla^2 \dot T= 0,\label{eq:t0eom}
     \end{eqnarray}
where $c_s^2 \equiv1/[3(1+R)]$ $R \equiv 3\rho_b/4\rho_\gamma$, $\beta \equiv 2/[45(1+R) \dot \tau]$, and we have ignored the effects of the gravitational potentials (which are small during radiation domination).  We will consider a simplified case (which has all the salient features of the full problem), in which the damping, $\beta$, is treated as constant in time and $c_{s}\simeq 1/\sqrt{3}$. In this case the equation of motion, Eq.~(\ref{eq:t0eom}), is  a damped, undriven, harmonic oscillator. We will model the effects of the spatial variation in the fine-structure constant by writing $\beta = \beta_0[1+ \varphi(\vec x)\partial \beta/\partial \varphi]$.

As in Ref.~\cite{He:2015msa}, we will solve the dynamical equation perturbatively, writing $T=T_0 + T_1$, where $T_1$ is first order in $\varphi$. Imposing the initial conditions $T_0(\vec k,0) = -\zeta(\vec k)/5$ [where $\zeta(\vec k)$ is the initial curvature perturbation] and $\dot T_0(\vec k, 0) = 0$ the zeroth-order solution is
\begin{eqnarray}
   T_0(\vec k, \eta) &=&  -e^{-\beta k^{2}\eta}\zeta(\vec k) /5\Bigg[\frac{k\beta}{\sqrt{k^2 \beta^2-c_s^2}}s(k,\eta)\\&+&c(k,\eta)\Bigg],\nonumber \\
   c(k,\eta)&\equiv&\cosh\left(k\eta\sqrt{k^{2}\beta^{2}-c_{s}^{2}}\right),\\
   s(k,\eta)&\equiv&\sinh\left(k\eta\sqrt{k^{2}\beta^{2}-c_{s}^{2}}\right),
   \label{eq:temp0}\end{eqnarray}
where the hyperbolic trigonometric functions take into account the transition between a strongly damped system when $k>c_s/\beta$ to a genuinely oscillatory system when $k<c_{s}/\beta$ via the identities $\cosh{(ix)}=\cos{x}$, $\sinh{(ix)}=i\sin{x}$.

The first order solution, $T_1$, satisfies the dynamical equation 
\begin{eqnarray}
 \ddot{{T}}_1 + c_s^2 k^2{T_1} + 2\beta k^2\dot{{T}}_1 = \mathcal{F}(\vec k,\eta),\label{eq:toy_alpha_eom}
 \end{eqnarray}
 where 
 \begin{align}
 \mathcal{F}(\vec k,\eta) \equiv 2\beta\frac{d \ln \dot{\tau}}{d\varphi} \int  \frac{d^3k_1}{(2\pi)^3} k_1^2 \dot{T}_0(\vec k_1,\eta)\varphi(\vec k_1 -\vec k),\label{eq:source}
 \end{align}
 so that $T_1$ behaves as a \emph{driven}, damped, harmonic oscillator. The Green's function of the left-hand side of Eq.~(\ref{eq:source}) is obtained by replacing the right-hand with $\delta(\eta-\zeta)$ and solving to obtain
 \begin{eqnarray}
 G(\eta-\zeta)=\left\{\begin{array}{ll}0&\mbox{if $\eta \leq \zeta$},\\\frac{s(k,\eta)}{k\sqrt{k^{2}\beta^{2}-c_{s}^{2}}}e^{-\beta k^{2}(\eta-\zeta)}&\mbox{if $\eta>\zeta$.}\end{array} \right.
 \end{eqnarray}The solution to Eq.~(\ref{eq:toy_alpha_eom}) is then obtained via the usual Green's function expression, $T_{1}(\vec{k},\eta)=\int_{0}^{\eta}d\zeta G(\eta-\zeta)\mathcal{F}(\vec{k},\zeta)$, yielding
 \begin{widetext}
 \begin{eqnarray}
   T_1(\vec k, \eta) &=& 2 \beta \frac{d \ln \dot{\tau}}{d\varphi} \int \frac{d^{3}k_{1}}{\left(2\pi\right)^{3}}\frac{  k_1^3 \varphi(\vec k - \vec k_1)\zeta(\vec k_1)  e^{-\beta  \eta  k_1^2}}{5 \left(k-k_1\right) \left(k+k_1\right) c_s^2} \Bigg\{k_1 e^{-\beta  \eta  \left(k^2-k_1^2\right)} \Bigg[2 \beta  c(k,\eta)-\frac{\left(c_s^2-2 \beta ^2 k^2\right) s(k,\eta)}{k \sqrt{\beta ^2 k^2-c_s^2}}\Bigg]\nonumber  \\&+&\frac{\left(c_s^2-2 \beta ^2 k_1^2\right) s(k_{1},\eta)}{\sqrt{\beta ^2 k_1^2-c_s^2}}-2 \beta  k_1 c(k_1,\eta)\Bigg\}.\label{eq:t1real}
 \end{eqnarray}
\end{widetext}
It is straightforward to then check that Eq.~(\ref{eq:t1real}) satisfies the necessary boundary conditions, $T_{1}(\vec{k},0)=0$ and $\dot{T}_{1}(\vec{k},0)=0$.

To obtain the SU approximation, we start with the unmodulated solution for the temperature perturbations in Eq.~(\ref{eq:temp0}), transform it to real space, and expand it in a power series in $\varphi$ so that 
\begin{eqnarray}
T^{\rm SU}(\vec x, \eta) &\simeq& T(\vec x,\eta; \varphi=0) + \varphi(\vec x) \frac{\partial T}{\partial \varphi},\\
&=& T_0(\vec x,\eta)+ T_1^{\rm SU}(\vec x,\eta).
\end{eqnarray}
\begin{widetext}where 
\begin{equation}
T_1^{\rm SU}(\vec{k},\eta)=\frac{ \beta}{\left(2\pi\right)^{3}}\frac{d\ln{\dot{\tau}}}{d\varphi}\int d^{3}k_{1}  \varphi(\vec{k}-\vec{k}_{1})\zeta(\vec{k_{1}})\frac{\beta  k_1 c_s^2 e^{-\beta  \eta  k_1^2} \left[\eta  k_1 \sqrt{\beta ^2 k_1^2-c_s^2} c(k_{1},\eta)-s(k_{1},\eta)\right]}{5 \left(\beta ^2 k_1^2-c_s^2\right)^{3/2}}.\label{eq:tsol_su}
\end{equation}
It is straightforward to verify that in the squeezed limit ($\vec{k}\to \vec{k}_{1}$), $T_1(\vec k, \eta)\to T^{\rm SU}(\vec k, \eta)$.

In a fixed realization of $\varphi$, correlations between observed modes are then induced after ensemble averaging over the primordial curvature fluctuation $\zeta$, and so evaluating the temperature perturbation at $\eta = \eta_*$ (the conformal time at the SLS)
\begin{eqnarray}
    \langle T(\vec{k})T^{*}(\vec{k}_{1})\rangle &\simeq& \left \langle T_{0}(\vec{k})T_{0}^{*}(\vec{k}_{1})\right\rangle+R(k,k_{1})\varphi(\vec{K}),\\
    \langle T_{0}(\vec{k})T_{0}^{*}(\vec{k}_{1})\rangle&=&P_{\rm TT}(k)\delta^{(3)}(\vec{k}-\vec{k}_{1}),\\P(k)&=&\frac{A}{25k^{3}}\Bigg[c(k,\eta)+\frac{k\beta}{\sqrt{k^2 \beta^2-c_s^2}}s(k,\eta)\Bigg]^{2},\\
    R(k,k_{1})&=&\frac{2 A\beta}{25} \frac{d \ln \dot{\tau}}{d\varphi} \frac{e^{-\beta  \eta  k_1^2}}{\left(k-k_1\right) \left(k+k_1\right) c_s^2} \Bigg\{k_1 e^{-\beta  \eta  \left(k^2-k_1^2\right)} \Bigg[2 \beta  c(k,\eta)-\frac{\left(c_s^2-2 \beta ^2 k^2\right) s(k,\eta)}{k \sqrt{\beta ^2 k^2-c_s^2}}\Bigg] \nonumber \\&+&\frac{\left(c_s^2-2 \beta ^2 k_1^2\right) s(k_{1},\eta)}{\sqrt{\beta ^2 k_1^2-c_s^2}}-2 \beta  k_1 c(k_1,\eta)\Bigg\}\Bigg[-\frac{k\beta}{\sqrt{k^2 \beta^2-c_s^2}}s(k_1,\eta)-c(k_1,\eta)\Bigg]+(\vec{k}\leftrightarrow \vec{k}_{1}).\label{eq:offdiarea}
\end{eqnarray}\end{widetext}where $\vec{K}=\vec{k}-\vec{k}_{1}$ and we have assumed a scale-invariant power spectrum of primordial curvature perturbations $P_{\zeta}(k)=A/k^{3}$. The interchange $\vec{k}\leftrightarrow \vec{k}_{1}$ indicates that to the first term of Eq.~(\ref{eq:offdiarea}) we must add the same term with the swap performed. These two first-order terms arise from the cross terms in the product $T(\vec{k})T^{*}(\vec{k}_{1})=\left[T_{0}(\vec{k})+T_{1}(\vec{k})\right]\left[T_{0}^{*}(\vec{k})+T_{1}^{*}(\vec{k})\right]$.

In the null hypothesis of no $\alpha$ fluctuations, only the zeroth-order solution $T_{0}$ contributes, while in the presence of a fixed $\alpha$ modulation with wave vector $\vec{K}$, the isotropy-breaking `response' $R(k,k_{1})$ function codifies the imprint of a fixed $\alpha$ fluctuation on off-diagonal correlations of the temperature.

The response may also be calculated using the separate-universe approximation [Eq.~(\ref{eq:tsol_su})], to obtain
\begin{widetext}
\begin{eqnarray}
R^{\rm SU} (k,k_{1})&=&\frac{A \beta }{25}\frac{d\ln{\dot{\tau}}}{d\varphi} \frac{\beta  c_s^2 e^{-\beta  \eta  k_1^2} \left[\eta  k_1 \sqrt{\beta ^2 k_1^2-c_s^2} c(k_{1},\eta)-s(k_{1},\eta)\right]}{k_{1}^{2} \left(\beta ^2 k_1^2-c_s^2\right)^{3/2}}\Bigg[-\frac{k\beta}{\sqrt{k^2 \beta^2-c_s^2}}s(k,\eta)-c(k,\eta)\Bigg]\nonumber\\&+&\vec{k}\leftrightarrow\vec{k}_{1}.
\end{eqnarray}
\end{widetext}
It is straightforward to verify that in the limit $\vec{k_{1}}\to\vec{k}$, which implies that $\vec{K}\to 0$, that $R^{\rm SU}(k,k_{1})=R(k,k_{1})$

\begin{figure}[ht]
\includegraphics[width=\columnwidth]{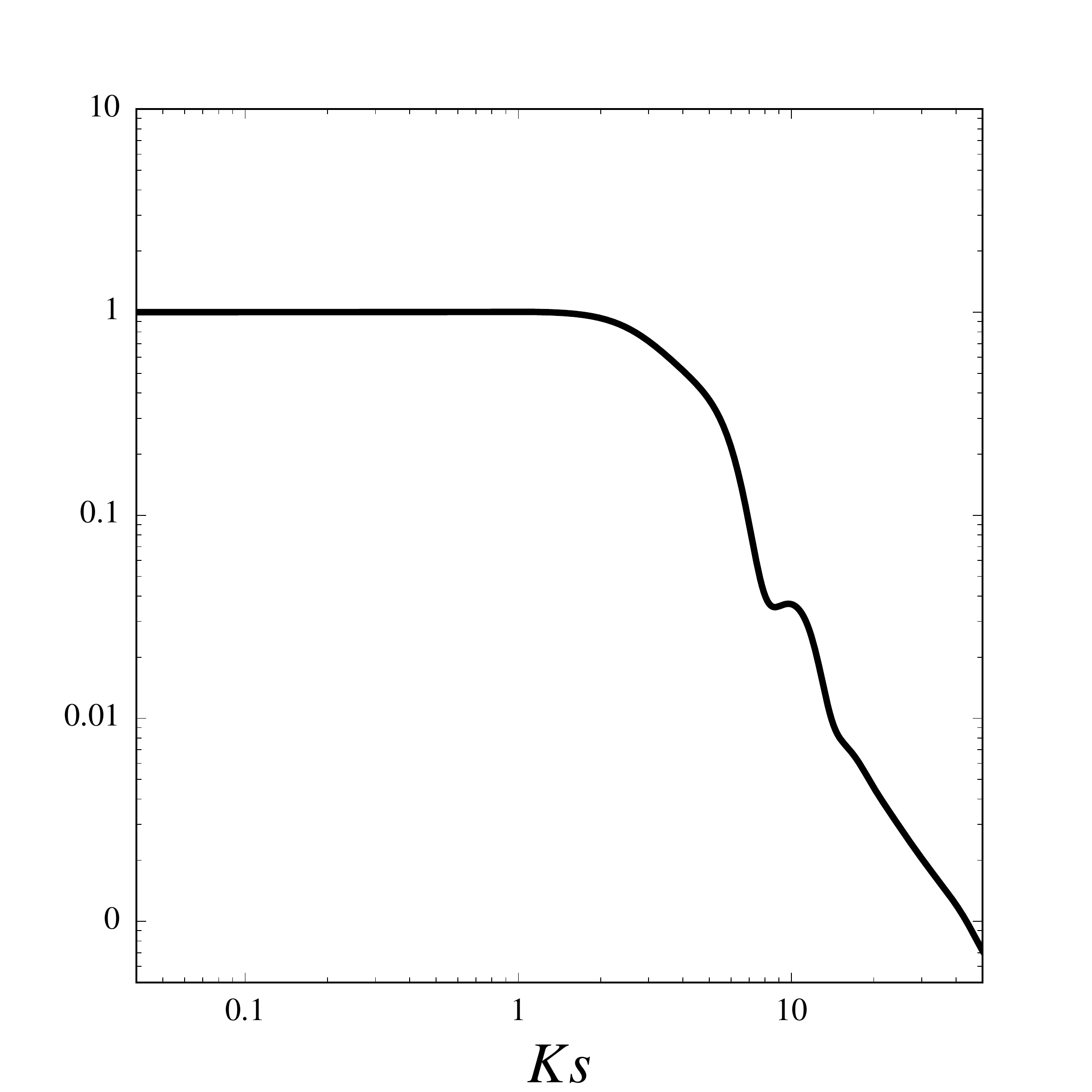}
\caption{Bias of separate-universe approximation minimum-variance estimator for $\varphi$ reconstruction. Horizontal axis is wave number of the $\alpha$-modulating mode in units of the inverse acoustic horizon.}
\label{fig:biasplot}
\end{figure} 

The analysis tools used to search for $\alpha$ variations in this paper [e.g.~the CMB off-diagonal correlations represented by Eq.~(\ref{eq:2pt})] are derived in the SU limit of a much more complete model, one that includes baryons, scattering terms, time-dependent gravitational potentials, neutrinos, and so forth. In this more complete model, a dynamical model would be much more challenging to obtain. For this toy model, where we have both SU and dynamical response functions, we can quantitatively assess how biased our inferences about the $\varphi$ field will be when the SU approximation is used.

To infer the Fourier transform $\varphi(\vec{K})$ of the modulating field from the `data' assuming the SU response, we may use the minimum-variance estimator
\begin{align}
\hat{\varphi}(\vec{K})=&N^{\rm SU}_{K}\int \frac{d^3 k}{\left(2\pi\right)^{3}} \frac{T(\vec{k})T(\vec{k}_{1})R^{\rm SU}(k,k_1)}{\tilde{P}_{\rm TT}(k)\tilde{P}_{\rm TT}(k_{1})},\\
\left(N_{K}^{\rm SU}\right)^{-1}=&\int \frac{d^{3}k}{\left(2\pi\right)^{3}}\frac{\left[R^{\rm SU}(k,k_1)\right]^{2}}{\tilde{P}_{\rm TT}(k)\tilde{P}_{\rm TT}(k_{1})},
\end{align} where the observed power spectrum
\begin{equation}
\tilde{P}_{\rm TT}(k)=P_{\rm TT}(k)+N_{\rm TT}(k)
\end{equation} includes the additional effect of a Poisson noise term, that is $N_{\rm TT}(k)=~{\rm const}$. 

The SU response does not perfectly reproduce the full dynamical response. If the dynamical response for some Fourier-space wave-vector triangle (a triplet $\vec{k},\vec{k}_{1}, \vec{K}$) is lower than the SU response, the absence of correlations for this triplet would lead to an erroneously low estimate $\hat{\varphi}(\vec{K})$ from this triangle, and vice versa. In our toy model, we can compute this bias
\begin{equation}
    b(K)\equiv \frac{\left \langle\hat{\varphi}(\vec{K})\right \rangle}{\varphi(\vec{K})}.\label{eq:bias}
\end{equation}A value $b=1$ indicates that $\varphi$ reconstruction based on the SU response is robust, while deviations indicate the limitations of this approximation.

In Ref.~\cite{He:2015msa}, this integral was evaluated in $3$ different ways: a fully analytic result valid in the $k \gg K $ limit, an evaluation in which the oscillatory (acoustic) features in the power spectrum are averaged out analytically prior to a numerical integral over the cosine $\vec{k}\cdot{\vec{K}}/(|\vec{k}||\vec{K}|)$, and a direct numerical evaluation of Eq.~(\ref{eq:bias}) including numerical noise. The response function here is much more complicated, and so we go directly to a fully numerical integral. 

We assume a Poisson-noise power spectrum with $1\%$ the amplitude of $P_{\rm TT}$ at $k= \pi/(\eta_* c_{s})$, commensurate with the $\sim 1\%$-level noise of modern CMB experiments. The noise term also regulates the effect of unphysical zeros in the power that occur in the toy model, but are `filled' in by the Doppler term and other effects in a more complete calculation. We check that once noise rises above a critical threshold (well below our chosen $N_{\rm TT}$), the resulting curves for $b(K)$ become independent of the noise level to $0.1\%$ accuracy. 

The result is shown in Fig.~\ref{fig:biasplot} (we use realistic values for the primordial baryon-photon plasma at decoupling: $\beta=2~{\rm Mpc}\ h^{-1}$ and $\eta_*=280~{\rm Mpc}\ h^{-1})$, with $K$ normalized relative to the acoustic horizon $s=\eta/\sqrt{3}$. We see that the minimum-variance estimator based on the SU approximation is unbiased for $K s\lesssim 2$, and is then biased as the SU approximation overestimates the response for the Fourier-space triangles dominating the estimate. This is only a toy model, and there  are  many  complicating  factors  that  could  change
the  result  at  the  order-unity  level.   This  onset  of  bias near the acoustic scale motivates us to proceed (conservatively), as was done in Ref.~\cite{He:2015msa}, and impose a cutoff of L= 100  in  our  trispectrum forecasts  and  lensing-based  reconstructions of the $\varphi$ power spectrum.

\section{Current constraints}
\label{sec:current}
To use \textit{Planck} data as a test for spatial variation of the fine-structure constant, we modified the publicly available Boltzmann solver \textsc{Camb}\footnote{\texttt{http://camb.info}} to compute the $\alpha$-modulated CMB power spectra and lensing-potential estimator given by Eq.~(\ref{eq:Alphacontrib}).  In particular, we modified \textsc{Camb} to compute the sum of the lensing-potential power spectrum and the $\alpha$-modulation contribution, given in Table.~II.  We compared these theoretical predictions to the \textit{Planck} data using the publicly available \textit{Planck} likelihood code \cite{Aghanim:2015xee} and the Markov Chain Monte Carlo (MCMC) code \texttt{cosmomc}\footnote{\texttt{http://cosmologist.info/cosmomc/}} \cite{Lewis:2002ah}.  

\begin{figure}[!ht]
\begin{center}
\resizebox{!}{8cm}{\includegraphics{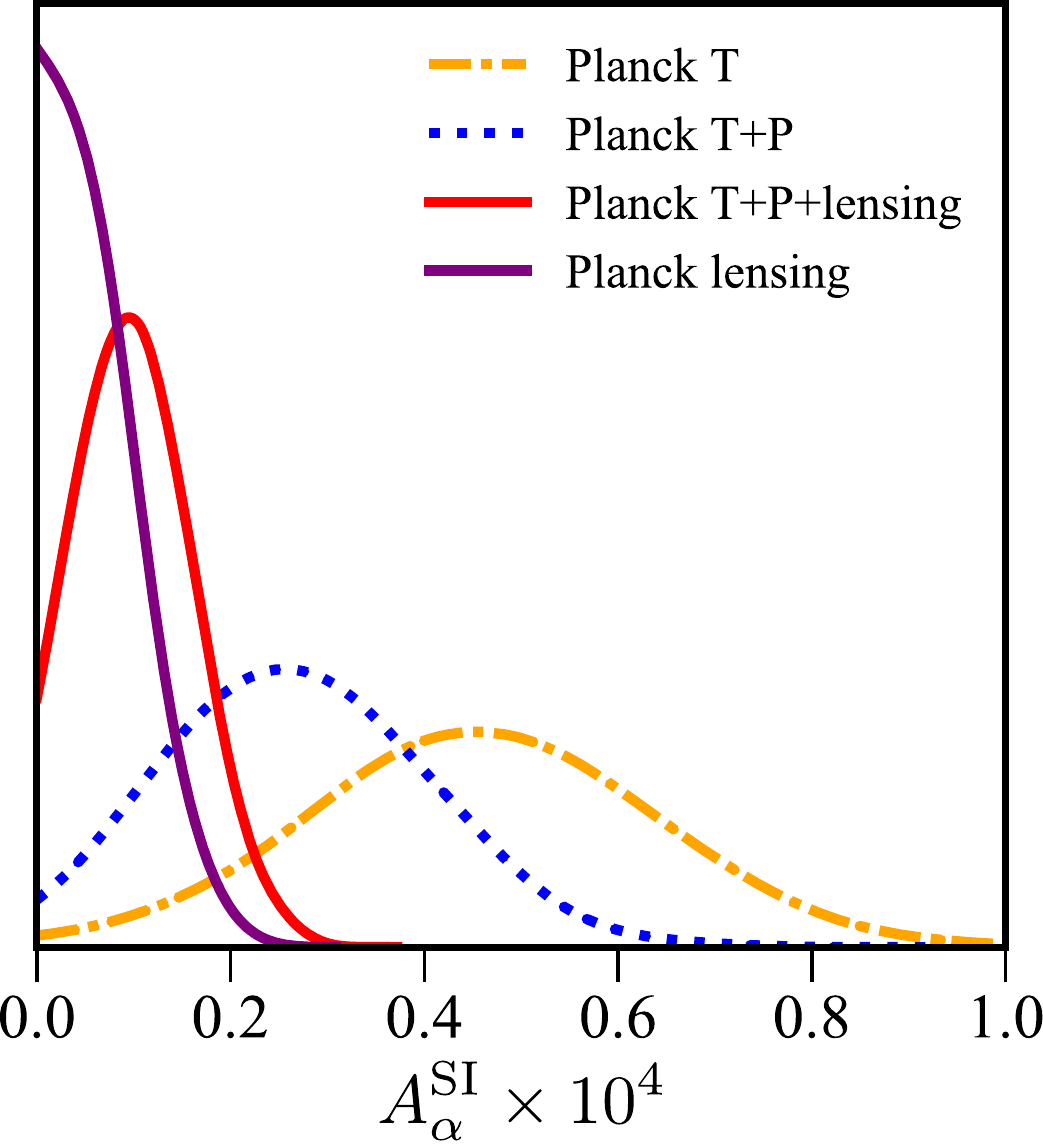}}
\caption{The 1D marginalized posterior for $A_\varphi$ using the three combinations of data sets discussed in the text. }
\label{fig:Avarphi1D}
\end{center}
\end{figure}

The \textit{Planck} data has been divided up into a large-angular-scale data set (low multipole number) and a small-angular-scale data set (high multipole number) \cite{Aghanim:2015xee}.  For all constraints we use the entire range of measurements for the TT power spectrum as well as the low multipole polarization (TE and EE) data, which we denote as `T+LowP'.  We also compute constraints using the entire multipole range of polarization measurements, denoted by `T+P'. The division between these two data sets is the multipole number $l =29$ which approximately corresponds to an angular scale of $\simeq 5^\circ$. In addition to the temperature and polarization power spectra we use the \textit{Planck} estimate of the lensing-potential power spectrum \cite{Ade:2015zua}.  We use the `aggressive' estimate of the lensing-potential power spectrum which extends down to $L_{\rm min} = 8$.   We used the \texttt{plik} likelihood \cite{Aghanim:2015xee} and varied all 27 \textit{Planck} nuisance parameters. 
\begin{table*}[hbtp!]
		\begin{tabular}{ l  c  c  c  c  }
			\hline
			\hline
			Parameter &  T+LowP & T+P & lensing & T+P+lensing    \\             
			\hline
			$\omega_b \dots\dots\dots$ & $0.02268\pm 0.00031$ & $0.02237\pm 0.00018$ & $0.02222\pm 0.00015$  & $0.02226\pm 0.00016$  \\
			$\omega_c \dots\dots\dots$ & $0.1156\pm 0.0027$ & $0.1185\pm 0.0016$ & $0.1193\pm 0.0014$  & $0.1190\pm 0.0014$  \\
			$n_s \dots\dots\dots$ & $0.9761\pm 0.0076$ & $0.9675\pm 0.0051$ & $0.9643\pm 0.0047$  & $0.9652\pm 0.0047$ \\
			$\log\left( 10^{10}\,A_s \right)$ & $3.045\pm 0.041$& $3.048\pm 0.040$ & $3.050\pm 0.024$  & $3.049\pm 0.025$ \\
			$\tau \dots\dots\dots..$ & $0.060\pm 0.021$ & $0.058\pm 0.020$ &  $0.059\pm 0.013$ &  $0.059\pm 0.014$ \\
			$H_0 \dots\dots\dots$ & $69.5\pm 1.3$ & $67.93\pm 0.74$ &  $67.46\pm 0.62$ & $67.60\pm 0.63$  \\
			$A_{\alpha}^{\rm SI}\times 10^4 \dots$ & $0.47\pm 0.18^{+0.35}_{-0.36}$ & $0.28^{+0.12\ +0.24}_{-0.15\ -0.26}$ & $< 0.16$ & $< 0.21$  \\
			\hline
		\end{tabular}
		\caption{Best-fit values and standard deviations for cosmological parameters with the three different \textit{Planck} data sets as described in the text. 
			All upper limits to $A_{\alpha}$ show 95\% C.~L.}
		\label{tab:constraints}
	\end{table*} 
\label{sec:constraints}

Our results are shown in Table \ref{tab:constraints}. The T + LowP data sets favor a non-zero $\alpha$ modulation with $A_{\alpha}^{\rm SI}=(4.7 \pm 1.8) \times 10^{-5}$. As demonstrated in Fig.~\ref{fig:alphaCls}, polarization data can break degeneracies present in a temperature-only analysis. Indeed, 
when we include the full polarization measurements from \textit{Planck} (i.e., `T+P') the best-fit value for the $\alpha$ power spectrum decreases to $A_{\alpha}^{\rm SI}=(2.7^{+1.2}_{-1.5}) \times 10^{-5}$. If we additionally include estimates of the lensing-potential power spectrum (i.e., `T+P+lensing'), the upper limit to the $\alpha$ modulation improves to $A_{\alpha}^{\rm SI}\leq 2\times 10^{-5}$ at 95\% C.~L. As noted in Secs. \ref{sec:eff} and Appendix \ref{sec:su_limits}, the response of the CMB to $\alpha$ fluctuations is not known for $\varphi$ multipole $L>100$, due to the breakdown of the SU approximation. 
For a conservative test of the varying $\alpha$ hypothesis, we also run MCMC chains with only the lensing power spectrum, and obtain the upper limit  $A_{\alpha}^{\rm SI}<1.6\times 10^{-5}$ at 95\% C.~L.

The 1D marginalized posterior on $A_{\alpha}^{\rm SI}$ using the different combinations of data sets is shown in Fig.~\ref{fig:Avarphi1D}. We also used $\hat{C}_{L}^{\phi\phi}$ to search for a white-noise power spectrum of $\alpha$ fluctuations, that is, $C_{L}^{\alpha}=A_{\alpha}^{\rm WN}$, and find that $A_{\alpha}^{\rm WN}\leq 2.3 \times 10^{-8}$ at 95\% C.~L.

\section{The optimal estimator}
\label{sec:optimal}
The constraints obtained in this work from the observed CMB trispectrum rely on the contribution of spatial fluctuations in $\alpha$ to the lensing-potential estimator. There is, however, an optimal $\varphi$ estimator which relies on the distinct (from lensing) off-diagonal CMB multipole correlations induced by the $\varphi$ field, as shown in Refs. \cite{Grin:2011tf,Grin:2011nk,He:2015msa} and summarized in Sec.~\ref{sec:est}. An analogous estimator was used to obtain the WMAP constraints to CIPs in Ref.~\cite{Grin:2013uya}. The Fisher information $F$ (which yields the minimum theoretically possible theoretical uncertainty in $A_{\alpha}$,~$\sigma_{A}=\sqrt{1/F}$) is given by
\begin{eqnarray}
F=\sum_{L}\frac{\left(2L+1\right)}{2}f_{\rm sky}\left(\frac{\partial C_{L}^{\Delta \Delta}}{\partial A_{\alpha}}   \right)^{2}\left(N_{L}^{\Delta \Delta}\right)^{-2}, \label{eq:fisher_optimal}
\end{eqnarray} 
where $N_{L}^{XX'}$ are defined in Ref.~\cite{Grin:2013uya} and computed under the null hypothesis. We use Eq.~(\ref{eq:fisher_optimal}) to forecast the sensitivity of \textit{Planck} and CMB-S4 (although \textit{Planck} data are public, we wish to compare our constraints to an optimal trispectrum analysis) to a scale-invariant angular power spectrum for $\varphi$.

\begin{figure}[!ht]
\begin{center}
\resizebox{!}{8cm}{\includegraphics{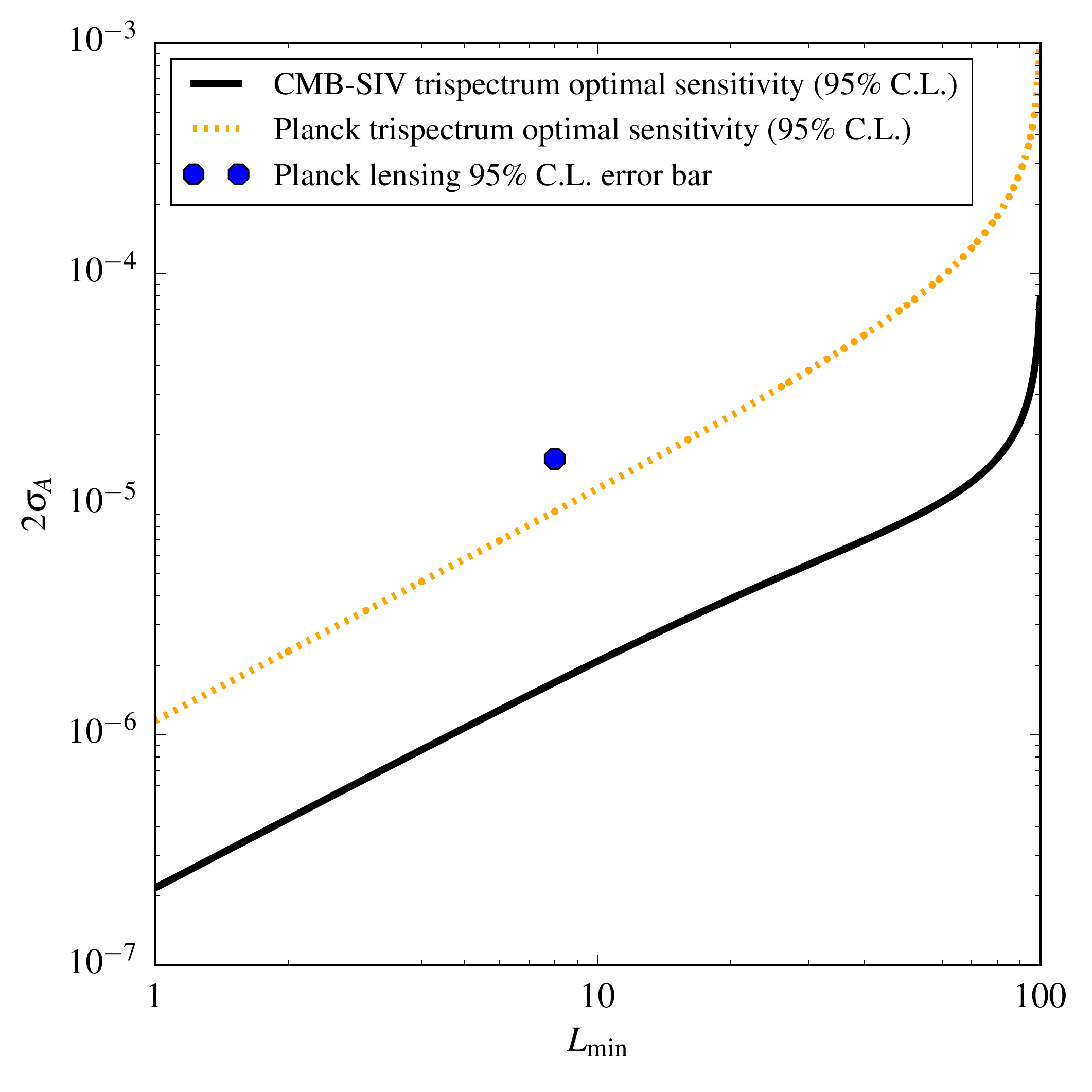}}
\caption{Sensitivity of optimal trispectrum-based estimator to a spectrum of scale invariant spectrum for $\varphi$. }
\label{fig:optimal}
\end{center}
\end{figure}

If we assume the null hypothesis, and that the posterior likelihood for $A_{\alpha}$ is Gaussian, we find that the optimum estimator has a $95\%$~C.~L. sensitivity of $A_{\alpha}^{\rm SI} \simeq 1.0 \times 10^{-5}$ for \textit{Planck} noise parameters, offering a slight improvement over the constraint from the $\alpha$ contribution to the lensing-potential estimator (see Fig.~\ref{fig:optimal} for an illustration as well as Sec.~\ref{sec:current}). We repeated this analysis [using Eq.~(\ref{eq:fisher_optimal})] with a shot-noise power-spectrum ($C^\alpha_{L}=A_\alpha^{\rm WN}$) and found that the optimal estimator has a $95\%$~C.~L. sensitivity of $A_\alpha^{\rm WN}\simeq
1.3 \times 10^{-8}$, again slightly lower than the limit obtained in Sec.~\ref{sec:current}. In other words, the constraints (obtained using the $\alpha$ contribution to the lensing-potential estimator) in Sec.~\ref{sec:constraints} are nearly optimal using \textit{Planck} data.

It is also interesting to consider the sensitivity of a future, nearly cosmic-variance limited (CVL) experiment, like the CMB-S4 concept \cite{Abazajian:2016yjj}. We use the noise parameters in Table \ref{table:plancksens}, and reconstruction noise as given in 
Eq.~(\ref{eq:oweight}), with the replacement $f_{l'Ll}^{\omega}\to h_{l'L l}^{\omega}$. We then use Eq.~(\ref{eq:fisher_optimal}) and find that CMB-S4 will be sensitive to scale-invariant $\alpha$ fluctuations with $A_{\alpha}^{\rm SI}\geq 1.9 \times 10^{-6}$ and $A_{\alpha}^{\rm WN} \geq 1.4 \times 10^{-9} $ (at $95\%$~C.~L. or greater). This difference, illustrated in Fig.~\ref{fig:optimal} for the scale-invariant case, is driven by the constraining power of a nearly CVL polarization experiment.

Given the fact that the trispectrum is so much more constraining than the $\varphi$-induced smoothing of the CMB power spectrum, we neglect primary power-spectrum constraints in this Fisher analysis. For futuristic experiments (like CMB-S4), the reconstruction noise for both lensing and $\alpha$ fluctuations may be low enough that lensing could introduce a significant bias \cite{Heinrich:2016gqe} to the estimators described in Sec.~\ref{sec:est}, requiring either a debiased minimum-variance estimator (as discussed in Ref.~\cite{Su:2011ff}) or a `delensed' CMB map (as discussed in Refs.~\cite{Smith:2008an,Smith:2010gu,Larsen:2016wpa}), in which lensing-induced correlations have been filtered out. We defer an analysis that includes these complications to future work, and simply note that Eq.~(\ref{eq:fisher_optimal}) quantifies the best $\alpha$ reconstruction we could achieve using the CMB.

\section{Discussion}
\label{sec:discussion}
We now explore if our results are consistent with $\alpha$ fluctuations being responsible for the anomalous smoothing of the CMB power spectra (e.g., Ref.~\cite{Aghanim:2018eyx}), or with the putative detection of an angular dipole in $\alpha$ seen in quasar spectra (e.g., Ref.~\cite{King:2012id}).

\subsection{The variation of the fine-structure constant and the anomalous smoothing of the CMB power spectrum} 

Weak gravitational lensing by clustered matter between us and the SLS causes two main effects on the CMB: it smooths the CMB power spectra (both temperature and polarization) and it generates correlations between different multipoles leading to a (non-Gaussian) connected part of the CMB trispectrum (see Ref.~\cite{Lewis:2006fu} and references therein). The CMB trispectrum can, in turn, be used to estimate the lensing potential power spectrum and the amplitude of this power spectrum predicts the level of smoothing of the CMB power spectra \cite{Calabrese:2008rt}. The internal consistancy of these two effects have been used to explore possible deviations from the standard cosmological model-- and any discrepancy may be due to physical processes which modulate the CMB anisotropies, such as spatial variations in $\alpha$. Recent CMB observations by the \textit{Planck} satellite have found that the level of smoothing of the CMB power spectra are 3 standard deviations larger than what is expected from the amplitude of the lensing potential power spectrum.  

\begin{figure}
\resizebox{!}{8.5cm}{\includegraphics{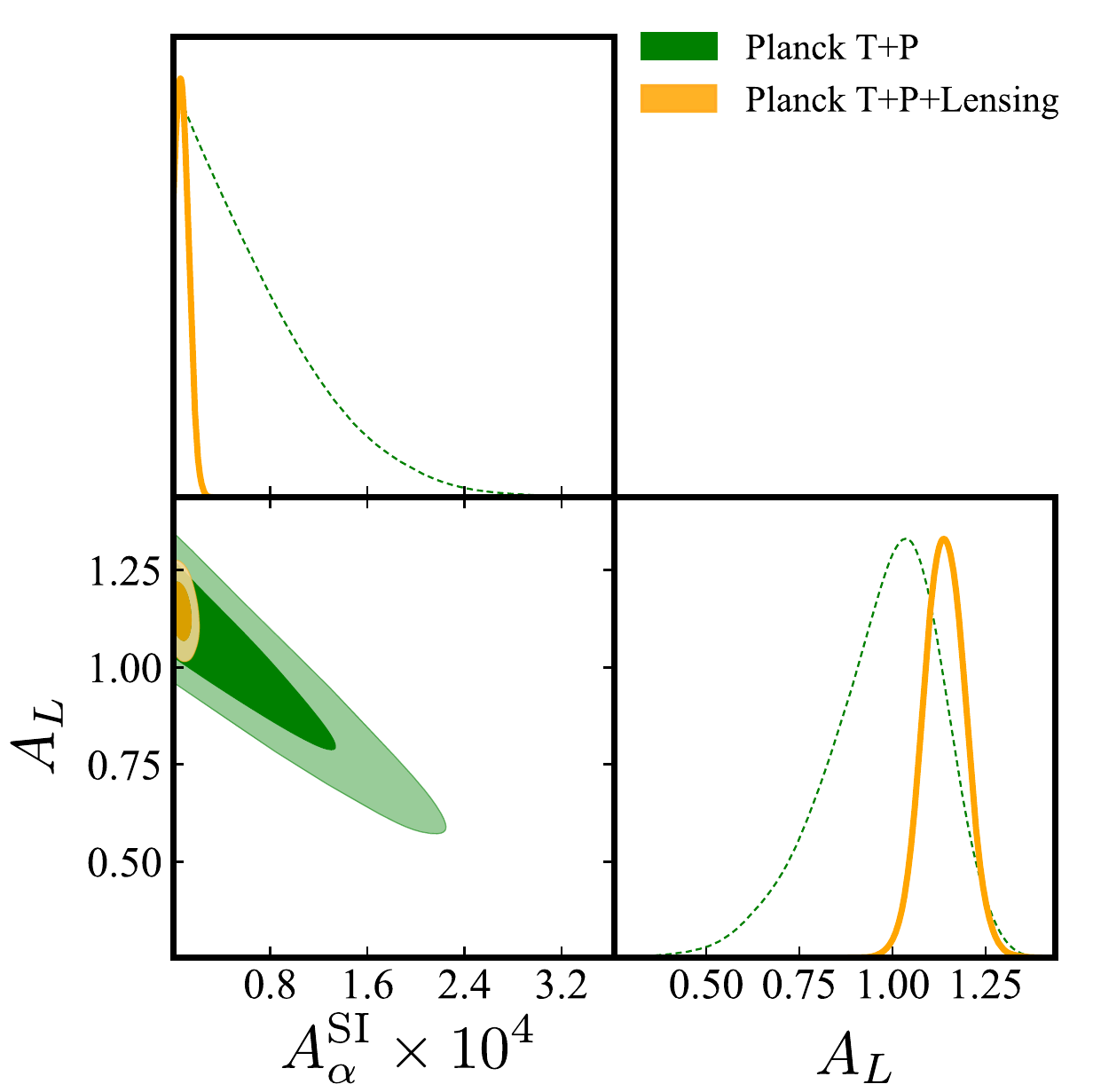}}
\caption{Constraints to $A_\alpha^{\rm SI}$ and $A_L$ from \textit{Planck} temperature (T), polarization (P), and estimates of the lensing pote tial power spectrum (Lensing). Without the lensing data, the degeneracy between $A_\alpha^{\rm SI}$ and $A_L$ causes $A_L$ to be consistant with its expected value of unity. Once we include the lensing data this degeneracy is broken and $A_L$ takes on an anomously large value.}
\label{fig:AL}
\end{figure} 

As shown in Fig.~\ref{fig:alphaClsRMS}, in the presence of a spatially varying $\alpha$ the CMB power spectra are smoothed. The observed anomalous smoothing of the CMB power spectra leads to the preference for a non-zero value of $A_\alpha^{\rm SI}$ when using the \textit{Planck} temperature and polarization data shown in Table \ref{tab:constraints}. 

In order to explore whether the additional modulation due to a spatially varying $\alpha$ might explain any additional smoothing we ran an MCMC with both $A_\alpha^{\rm SI}$ and $A_L$, where $A_L$ is a parameter that controls the level of the smoothing of the power spectra due to weak lensing (and \emph{does not} affect the amplitude of the lensing potential power spectrum) and has an expected value of $A_L = 1$ in the standard cosmological model. Fig.~\ref{fig:AL} shows that with just the temperature and polarization (T+P) data there is a degeneracy between these two amplitudes and they must take on values $A_\alpha^{\rm SI} < 1.8 \times 10^{-4}$ at 95\% C.~L.~and $A_L = 0.98^{+0.18\ +0.28}_{-0.11\ -0.32}$-- so that $A_L$ is fully consistent with unity. 

When we include the estimates of the lensing potential power spectrum (T+P+Lensing) the value of $A_\alpha^{\rm SI}$ is much more constrained ($A_\alpha^{\rm SI} < 1.7 \times 10^{-5}$ at 95\% C.~L.) and is unable to account for the anomalous smoothing of the CMB power spectra so that $A_L =1.14 \pm 0.056$- about 2.5 standard deviations larger than its expected value. 

This shows that the anomolous smoothing of the CMB power spectra is unlikely to be explained by a spatial variation of $\alpha$.

\subsection{The consequences for dynamical models of fine-structure constant variation}

With limits in hand it is interesting to explore the implications of our constraints for specific dynamical $\alpha$ models. This will allow us to propagate our constraints at the CMB to late times in order to compare to the putative measurement of the angular dipole in $\alpha$ from quasar spectra. 

In one scenario \cite{Sigurdson:2003pd}, $\alpha$ fluctuations are sourced by a scalar field with low (but non-negligible) mass $m_{\phi}$, with quadratic couplings to standard-model gauge fields.\footnote{In a full high-energy theory of varying $\alpha$, large quantum corrections to $m_{\phi}$ would result. We consider this scenario to be an effective theory and neglect quantum corrections.} It is interesting to ask if this model can simultaneously stay within our constraints to $(\sigma_{\alpha}/\alpha_{0})$, but explain the claimed dipole of $\alpha$ in analyses of QSO spectra \cite{Webb:2010hc}. To answer this question, we must consider the time evolution of $\varphi$, given the vastly disparate redshifts of the two observables involved.

The scalar field equation of motion is
\begin{equation}
\ddot{\phi}+3H\dot{\phi}+\left(m_{\phi}^{2}+\frac{\eta}{t^{2}}\right)\phi=0.\end{equation} Here $\eta$ is proportional to the fraction of the total matter density that contributes to the expectation value $\left \langle E^{2}-B^{2}\right \rangle$ for the electromagnetic-field Lagrangian density and the coupling of $\phi$ to matter, $H$ is the usual Hubble parameter, and $m_{\phi}$ is the mass of the light field, while $E$ and $B$ are the electric and magnetic field. When $\eta$ is negligible, the background field evolves according to $\phi_{0}(t)\propto \sin{(m_{\phi}t)}/({m_{\phi}t})$, as do perturbations $\delta \phi(\vec{x},t)$, so long as they are still outside the horizon. 

In this model, the usual Maxwell Lagrangian ($F_{\mu \nu}F^{\mu \nu}$) acquires an additive correction that scales as $\phi^{2}F_{\mu \nu}F^{\mu \nu}$. When canonical field normalization is imposed on the photon ($A_{\mu}$), we see that a spatially varying $\alpha$ arises, with the lowest-order contribution from spatial fluctuations given by $\delta \alpha/\alpha=\varphi\propto \phi_{0}(t)\delta \phi(t)$. For particle masses $m_{\phi}\gtrsim H_{\rm rec}\sim 10^{-28}~{\rm eV}$, the power spectrum (and variance) of $\alpha$ thus evolves as $t^{-4}\propto (1+z)^{6}$, where the last scaling emerges because most of the redshift ($z$) interval between decoupling and today occurs primarily during matter domination. In related models, this scalar can even be a significant component of the dark matter \cite{Stadnik:2015kia}.

Evolving the 95\%~C.~L. limits to a scale-invariant $\varphi$ spectrum from Planck and forward in time to $z\sim 2$ yields an r.m.s.~dipole value of \begin{equation} \sqrt{L(L+1)\left.C_{L}^{\alpha}\right|_{L=1}/(2\pi)}\sim 10^{-12},\end{equation} many orders of magnitude below the $\alpha $ dipole inferred in Ref.~\cite{Webb:2010hc} ($0.97\pm 0.88 \times 10^{-5}$ at 95\%~C.L.).\footnote{Note that these limits are much more stringent than those estimated in Ref.~\cite{OBryan:2013nip}, where it is assumed that $\delta \alpha/\alpha_{0} \propto \delta \phi(t)\sim 1/t$ rather than $\delta \alpha/\alpha_{0}\sim 1/t^{2}$.} In other words, the model of Ref.~\cite{Sigurdson:2003pd} cannot simultaneously accommodate CMB measurements and hints from QSO spectra of a spatial dipole in $\alpha$.

In contrast, the \textit{total} electromagnetic Lagrangian of the BSBM theory \cite{Bekenstein:1982eu,Barrow:2002db} is $\propto e^{-2\phi}F_{\mu \nu}F^{\mu \nu}$, with a homogeneous scalar field equation of motion of the form \begin{equation}\ddot \phi +3H\dot{\phi}\propto \rho_{m}e^{-2\phi}.\end{equation} The resulting super-horizon spatial variations in $\alpha$ behave as $\delta \alpha/\alpha_{0}=2\delta \phi={\rm const}$ during matter domination, while sub-horizon fluctuations grow \cite{Barrow:2002zh}. As a result, there is negligible decay in the r.m.s.~fluctuation $\alpha$ fluctuations on the scales of interest, and existing limits from the CMB [ $\left(\sigma_\alpha/\alpha_{0}\right)_{\theta>10^{\circ}}\sim 10^{-3}$] do not rule out a putative QSO dipole of $\sigma_{\alpha}/\alpha_{0}\simeq  10^{-5}$ for this theory. Thus, even a futuristic experiment like CMB-S4 would be unable to rule out the BSBM explanation for the QSO dipole.

Another interesting possibility is the `runaway dilaton' model \cite{Damour:2002nv,Damour:2002mi}. Light scalars (dilatons) controlling the volume of extra dimensions appear in some variants of string theory. To prevent dilatons from causing highly constrained violations of the weak equivalence principle, one can posit a large dilaton mass, \textit{or} rely on the matter couplings of the dilaton ($\propto e^{-\phi}$) and a non-canonical kinetic term to dynamically drive it towards weak coupling via cosmological evolution \cite{Damour:1994zq}. The latter option is the `runaway dilaton' scenario. 

One interesting feature of this model is that the amplitude of spatio-temporal $\alpha$ fluctuations is related to the amplitude of primordial density fluctuations (and thus $A_{s}$) \cite{Damour:1994zq,Damour:2002nv}. Spatial fluctuations evolve as $\varphi\propto \ln{(1+z)}$ in this model \cite{Damour:2002nv,Martins:2017yxk}. With this scaling, a scale-invariant spectrum of $\alpha$ fluctuations saturating our CMB limits would decay to an r.m.s.~$\alpha$ dipole of $\sim 2 \times 10^{-4}$ at $z\sim 2$. At CMB-S4 sensitivity levels, however, this would improve to an r.m.s.~$\alpha$ dipole of $9 \times 10^{-5}$, closer to the QSO hints of Ref.~\cite{Webb:2010hc}. 

As $\alpha$ fluctuations are correlated with primordial density fluctuations, perhaps an additional improvement in sensitivity could be achieved by correlating $\alpha$ fluctuations with CMB observables. This would allow us to use the observed bispectrum (rather than the trispectrum) to search for variations in the fine-structure constant. Furthermore, all the estimates in this section compare an r.m.s.~dipole signal to the observed QSO dipole. A more rigorous analysis could (in the context of a specific models and its equations of motion for $\alpha$ perturbations) map this dipole to a predicted pattern of isotropy breaking in CMB maps, perhaps improving sensitivity, even bringing a runaway-dilaton explanation for the QSO results to be empirically tested using \textit{Planck} and other data. In any of these models, depending on the details, the scalar field could be the cosmological dark energy or just an unrelated scalar; in either case, the time evolution of $\phi_{0}$ and $\delta \phi$ is related to the evolution of the dark-energy density in a predictable way \cite{Martins:2017yxk}.

\section{Conclusions}
\label{sec:conclude}
A variety of theoretical ideas motivate the consideration of a spatially-varying fine-structure constant. By modulating the recombination history and Thomson scattering rate of the early universe, such a scenario would alter CMB statistics. The mathematical formalism is similar to that used in studies of compensated isocurvature perturbations (CIPs) and weak gravitational lensing of the CMB, but with a specific response to the physics of $\alpha$ modulation. Using a toy model, we find that this response falls off for scales $L\geq 100$, just as for CIPs.

Here, we used measurements of CMB trispectra (as captured by the optimal estimator of the weak-lensing power spectrum) and power spectra to test for the presence of a scale-invariant power spectrum of $\alpha$ fluctuations. This is an interesting possibility as any fluctuation in $\alpha$ sourced by a massless field present during inflation would naturally have a scale-invariant spectrum.  Using just the $\alpha$ contribution to the lensing potential power spectrum for a scale invariant power spectrum ($C_L^{\alpha} = A_\alpha^{\rm SI}/[L(L+1)]$), we find the constraint (at 95\%~C.~L.) $A_\alpha^{\rm SI} < 1.6 \times 10^{-5}$, which implies a fractional variation in $\alpha$ on tens of degrees or larger of $(\sigma_\alpha/\alpha_0)_{\theta > 10^\circ} < 2.5 \times 10^{-3}$ [constraints to the variance, a derived parameter, are obtained using Eq.~(\ref{eq:normps}) but with the multipole range $2\leq L\leq 20$]. For a constant (white noise) power spectrum $C_{L}=A_\alpha^{\rm WN}$, we find that $A_\alpha^{\rm WN} < 2.3 \times 10^{-8}$ and $ (\sigma_\alpha/\alpha_0)_{\theta > 10^\circ} < 8.9 \times 10^{-4}$, all at $95\%$~C.~L. This is an improvement over the constraints found using the 2013 \textit{Planck} data \cite{OBryan:2013nip}.

Furthermore, we find that at \textit{Planck} noise levels, the sensitivity of our estimator (based on lensing data products) is nearly optimal, as shown by the Fisher analysis for scale-invariant $\alpha$ fluctuations in Sec.~\ref{sec:optimal}; we performed the same analysis for white-noise power spectrum, and again found that our lensing-based constraint is comparable in sensitivity to a full trispectrum analysis for \textit{Planck} noise levels.

Future experiments (e.g.~CMB-S4 \cite{Abazajian:2016yjj}) will achieve nearly cosmic-variance limited measurements of CMB polarization, thus pushing the sensitivity to scale-invariant $\alpha$ fluctuations as low as $A_\alpha^{\rm SI}= 1.9 \times 10^{-6}$ (or in the white-noise case, $A_\alpha^{\rm WN}=1.4 \times 10^{-9}$), or variances as low as $(\sigma_\alpha/\alpha_0)_{\theta > 10^\circ} = 8.6 \times 10^{-4}$ (or in the white-noise case, $(\sigma_\alpha/\alpha_0)_{\theta > 10^\circ} = 2.2 \times 10^{-4}$). 

We considered the possibility that $\alpha$ may alleviate the anomalously large smoothing of the CMB power spectra relative to the amplitude of the lensing potential power spectrum. In the end, scale-invariant $\alpha$ fluctuations cannot resolve this tension, due to trispectrum estimates of the lensing potential power spectrum. This conclusion depends on the shape of the modulating field's power spectrum and precise response of observables to the modulating field. In future work, it will be interesting to explore what type of long-wavelength modulation could explain anomalous smoothing of the CMB power spectra while satisfying trispectrum constraints. 

Above, we used our phenomenological limits to estimate  constraints to actual dynamical theories of varying $\alpha$, appropriating a CMB analysis that treated $\alpha$ as spatially varying but constant in time. We thus remind the reader that the translation of our constraints to limits on specific models of varying $\alpha$ are just order-of-magnitude estimates. Robust tests require a proper evolution of the background $\alpha$ value, a proper relativistic treatment of perturbation evolution, and a computation of the imprint of these dynamics on observables using a Boltzmann code like \textsc{Camb} \cite{camb} or \textsc{Class} \cite{Lesgourgues:2011re}, with appropriate modifications for the model of interest.

Additional improvements could also follow from analyzing CMB trispectra directly (rather than a lensing power-spectrum based estimator) and doing a map-level analysis for the time-evolved imprint of the claimed QSO dipole. We will pursue a more complete analysis along these lines in future work, which will also update our analysis to include power spectra and lensing \cite{Aghanim:2018oex} results from the \textit{Planck} 2018 data release \cite{Aghanim:2018eyx}, which indeed contain statistically marginal hints for CIPs, which could also be caused by $\alpha$ fluctuations \cite{Akrami:2018odb}. Indeed, as noted in Ref.~\cite{Hill:2018ypf}, there are still systematic (but unsubtracted) biases contributing to estimators of  non-Gaussianity in the CMB. These could also affect observable signatures of $\alpha$ modulation; a full trispectrum analysis including these biases and a variety of interesting theoretical possibilities (CIPs, $\alpha$ fluctuations, etc...) is thus in order.

Looking beyond CMB anisotropies, the full network of bound-bound and bound-free transition during the recombination era will produce spectral distortions of the CMB away from a perfect thermal spectrum (See Refs.~\cite{RubinoMartin:2006ug,Chluba:2015gta} and references therein). The rates of the relevant transitions depend very sensitively on $\alpha$, and so a futuristic measurement of spatially-dependent CMB spectral distortions from recombination lines would offer an interesting (and more primordial) test of the possibilities explored here. Furthermore, the rate of diffusion damping /efficiency of generating CMB spectral distortions all depend sensitively on $\alpha$ \cite{Chluba:2012gq,Khatri:2013dha}. Anisotropies of continuum CMB spectral distortions could thus also be an interesting test of spatially variations in $\alpha$ (as well as to time evolution of the background value, as noted in Ref.~\cite{Hart:2017ndk}).

In coming decades, observations of absorption in the 21-cm (hyperfine) transition of neutral hydrogen may help us to finally understand the `dark ages', the epoch between CMB decoupling at $z\sim 1090$ and the formation of the first stars near $z\sim 10-20$ (see Ref.~\cite{2012RPPh...75h6901P} and references therein for a more comprehensive discussion). As noted in Refs. \cite{Khatri:2007yv}, the $21$-cm line rest frame frequency scales as $\nu_{21}\propto \alpha^{4}$, the Einstein rate coefficient for the relevant decay scales as $A\propto \alpha^{13}$, and the spin-changing collisional cross sections of hydrogen also depend sensitively on $\alpha$. As a result, $21$-cm cosmology should provide a new probe of spatial fluctuations in $\alpha$, with the added advantage that measurements (by experimental efforts like HERA \cite{DeBoer:2016tnn} and SKA \cite{Maartens:2015mra}) at many redshifts should facilitate stringent tests of the time evolution of perturbations in different models of spatially varying $\alpha$. 

\acknowledgments We thank the Provosts' Offices of Swarthmore and Haverford Colleges for supporting this research. We thank Jens Chluba, Marc Kamionkowski, Wayne Hu, Yevgeny Stadnik, Victor Flambaum, Asantha Cooray, and John O'Bryan for stimulating discussions.  We thank Yacine Ali-Ha\"{i}moud for thorough comments on this manuscript and stimulating discussions. Part of this work was completed at the Aspen Center for Physics, which is supported by National Science Foundation grant PHY-1607611. This work was supported in part by the National Science Foundation under Grant No. NSF PHY-1125915 at the Kavli Institute for Theoretical Physics (KITP) at UC Santa Barbara. DG thanks KITP for its hospitality during the completion of this work.  

\begin{appendix}

\section{Detailed expressions for the $\alpha$-induced CMB off diagonal correlations}
\label{sec:details}
\begin{table*}[!ht]
\begin{tabular*}{11cm}{@{\extracolsep{\fill}}cccc}
\\\noalign{\smallskip}\hline\noalign{\smallskip}
{\rm XX}$'$&$f^{\rm X X'}_{l Ll'}$&$h^{\rm X X'}_{l Ll'}$&$l+l'+L$\\\noalign{\smallskip}\hline\noalign{\smallskip}
TT &$ \tilde{C}_{l}^{\rm TT}{\,}_0F_{l' L l}+\tilde{C}_{l'}^{\rm TT}{\,}_0F_{l L l'}$&$ \left(\tilde{C}_{l}^{ \rm T,dT}+\tilde{C}_{l'}^{\rm T,dT}\right){\,}_0H_{l L l'}$&even \\\noalign{\smallskip}
TE&$ \tilde{C}_{l}^{ \rm TE}{\,}_2F_{l' L l}+\tilde{C}_{l'}^{ \rm TE}{\,}_0F_{l L l'}$&$ \tilde{C}_{l}^{\rm T,dE}{\,}_2H_{l L l'}+\tilde{C}_{l'}^{\rm  E,dT}{\,}_0H_{l L l'}$& even  \\ \noalign{\smallskip}
TB&$i\tilde{C}_{l}^{ \rm TE}{\,}_2F_{l' L l} $&$i\tilde{C}_{l}^{\rm T,dE}{\,}_2H_{l L l'} $& odd
\\ \noalign{\smallskip}
EE&$ \tilde{C}_{l}^{ \rm EE}{\,}_2F_{l' L l}+\tilde{C}_{l'}^{ \rm EE}{\,}_2F_{l L l'}$&$ \left(\tilde{C}_{l}^{ \rm E,dE}+\tilde{C}_{l'}^{ \rm E,dE}\right){\,}_2H_{l L l'}$& even  \\ \noalign{\smallskip}
EB &$ i\left[\tilde{C}_{l}^{\rm EE}{\,}_2F_{l' L l} -\tilde{C}_{l'}^{ \rm BB}{\,}_2F_{l L l'}\right]$&$ i\left(\tilde{C}_{l}^{ \rm E,dE}{\,}+\tilde{C}_{l'}^{\rm B,dB}\right){\,}_2H_{l L l'}$& odd   \\ \noalign{\smallskip}
BB&$ \tilde{C}_{l}^{ \rm BB}{\,}_2F_{l' L l}+\tilde{C}_{l'}^{\rm BB}{\,}_2F_{l L l'}$& $(\tilde{C}_{l}^{\rm B,dB}+ \tilde{C}_{l'}^{\rm B,dB}){\,}_2H_{l L l'}$ & even  \\\noalign{\smallskip}\hline\noalign{\smallskip}\end{tabular*}
\caption{The lensing and $\alpha$-modulation response functions. ``Even'' and ``odd'' indicate that the functions are non-zero only when $L + l + l'$ is even or odd, respectively. To translate from the conventions of Ref.~\cite{He:2015msa} we need to swap $l \leftrightarrow l'$ which leads to a minus sign for the two odd responses, $EB$ and $TB$.  Note that the B-mode autocorrelation, BB, vanishes at linear order in the $\varphi$ field.} 
\label{tab:CIP_resp}
\end{table*}

Deflections of CMB photons and higher-order modulations of the transfer functions produce off-diagonal CMB correlations for fixed lens and $\varphi$ realizations, given by \cite{Okamoto:2003zw,Grin:2011nk,Grin:2011tf}
\begin{eqnarray}
\VEV{X_{l m} X'_{l'm'}}&\big|_{\rm lens,\alpha}& = \tilde{C}_l^{X X'} \delta_{l l'} \delta_{m -m'} (-1)^m\nonumber \\ &+& 
\sum_{LM} (-1)^M \wigner l m {l'} {m'} L {-M} \nonumber \\ &\times&\bigg[\phi_{LM} f^{XX'}_{l Ll'}+ \varphi_{LM} h^{XX'}_{l L l'}\bigg],
\label{eq:2pt}
\end{eqnarray}
 where $f^{XX'}_{l L l'}$ and $h^{XX'}_{l L l'}$  are the lensing/$\alpha$ response functions for different quadratic pairs (see Table \ref{tab:CIP_resp}) and are defined in terms of the unmodulated power spectrum, $\tilde{C}_l^{XX'}$, the appropriately weighted Wigner coefficients,
\begin{eqnarray}
_{\pm s} G_{l L l'} &\equiv& \left[L(L+1) + l'(l'+1) - l (l + 1)\right]\nonumber \\&\times& \sqrt{\frac{(2L+1)(2 l +1)(2 l'+1)}{16\pi}} \left(\begin{array}{ccc}l & L & l' \\ \pm s & 0 & \mp s\end{array}\right),\nonumber \\ \\
_{\pm s}H_{l L l'} &\equiv& \sqrt{\frac{(2L+1)(2 l +1)(2 l'+1)}{4\pi}} \left(\begin{array}{ccc}l & L & l' \\ \pm s & 0 & \mp s\end{array}\right), \nonumber \\
\end{eqnarray}
and the derivative power spectra \begin{equation}
C_l^{\rm X,dX'} \equiv \frac{2}{\pi} \int k^2 dk P_{\zeta}(k) X_{l}(k)\frac{dX'_l(k)}{d\varphi}.\label{eq:estnumdir} \end{equation}
The multipole moments of the lensing-potential are denoted by $\phi_{LM}$. This formalism was first developed for $\alpha$ fluctuations in Ref.~\cite{OBryan:2013nip}. Here we use the equivalent notation of Ref.~\cite{Smith:2017ndr}.

Under the null hypothesis (i.e., no $\alpha$ variation), the minimum-variance estimator for the `deflection field'
$d_{LM}^{\omega}\equiv\sqrt{L(L+1)}\phi_{LM}$ from a single pair $\omega=XX'$ of observables is \cite{Okamoto:2003zw,Namikawa:2011cs},
\be
\hat d^{\omega}_{LM} = A_L^\omega \sum_{l m, l'm'} \left(-1\right)^{M} X_{l m} X'_{l' m'} \wigner l m {l'} {m'} L {-M} g^{\omega}_{l l'L},
\label{eq:dhat}
\ee
where $A_L^\omega$ and $g^\omega_{l l'L}$ are 
\begin{eqnarray}
A_L^{\omega} =& L(L+1)(2L+1)\left\{\sum_{l_1 l_2} g^\omega_{l_1 L l_2} f^\omega_{l_1 L l_2}\right\}^{-1},\label{eq:norm}\\
g^{\omega}_{l Ll'} \equiv&\frac{C^{XX,{\rm t}}_{l'} C^{X'X',{\rm t}}_l f^{\omega*}_{l L l'} - (-1)^{l + L + l'} C_{l}^{XX',{\rm t}} C^{XX',{\rm t}}_{l'} f^{\omega*}_{l' L l}}{C_{l}^{XX,{\rm t}} C_{l'}^{XX,{\rm t}} C_{l}^{X'X',{\rm t}}C_{l'}^{X'X',{\rm t}} - (C_{l}^{XX',{\rm t}}C_{l'}^{XX',{\rm t}})^2}.\label{eq:optimal_filter}
\end{eqnarray} 

From this we can construct an optimal estimator for the lensing-potential power spectrum
\be
\hat C_L^{\phi \phi} = \frac{1}{2L+1} \sum_{\omega,\beta}\sum_{M=-L}^L v^\omega_{L}v^\beta_{L}  \frac{\hat{d}^\omega_{LM}\hat{d}^{*\beta}_{LM}}{L(L+1)}-B_L,
\label{eq:CIPestp}
\ee
where the optimal weights for the minimum-variance estimator are given by 
\cite{Okamoto:2003zw}
\begin{eqnarray}
v^\omega_{L} &\equiv& N_L^{\rm mv} \sum_{\beta} ({\bf N}_L^{-1})^{\omega \beta}, \\
N_L^{\omega \beta} &\equiv& \frac{A_L^{\omega *} A_L^\beta}{L(L+1)(2L+1)} \sum_{l_1 l_2} \big\{g_{l_1 L l_2}^{\omega*} \left[C^{\rm XY,{\rm t}}_{l_1} C^{\rm X'Y',{\rm t}}_{l_2} g^\beta_{l_1Ll_2}\right.\nonumber \\&+&\left.(-1)^{L+l_1+l_2} C^{\rm XY',{\rm t}}_{l_1} C^{\rm X'Y,{\rm t}}_{l_2} g^\beta_{l_2 L l_1}\right]\big\}, \quad\label{eq:oweight} \\
N_L^{\rm mv} &\equiv& \left[\sum_{\omega \beta} ({\bf N}_L^{-1})^{\omega \beta}\right]^{-1}.
\end{eqnarray} 

Using the same construction, we can form an optimal estimator for $C_L^{\alpha}$ by replacing $f^\omega_{l'Ll}$ with $h^\omega_{l'Ll}$ in Eqns.~(\ref{eq:norm}) and (\ref{eq:optimal_filter}).

\section{Power spectra derivatives}
\label{sec:step_optimal}

\begin{figure*}
\resizebox{!}{10cm}{\includegraphics{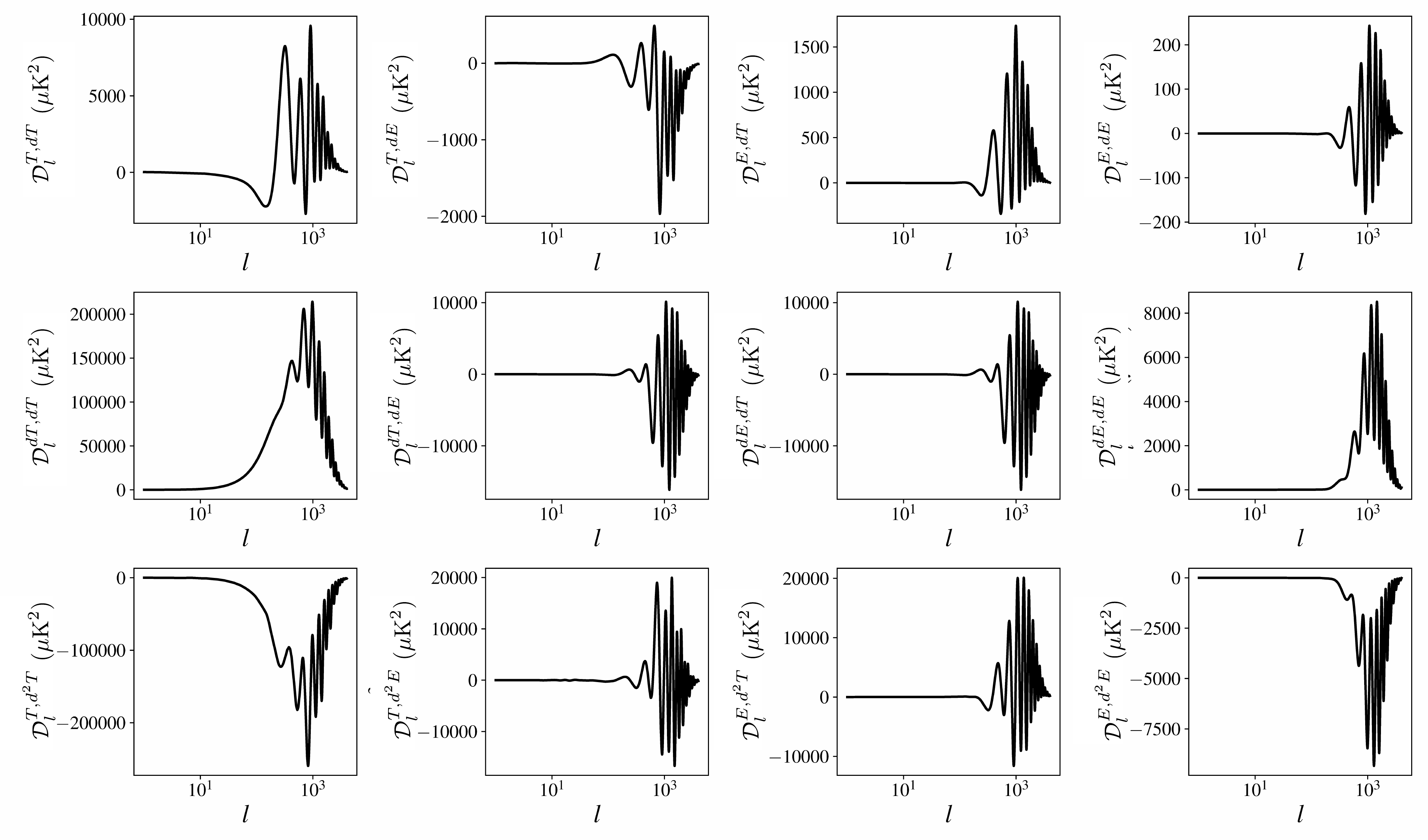}}
\caption{The derivative power spectra used in this paper, which agree (up to changes in fiducial cosmological parameters) with those in Refs. \cite{Sigurdson:2003pd} and \cite{OBryan:2013nip}.}
\label{fig:Clderivs}
\end{figure*} 

\begin{figure}
\resizebox{!}{5cm}{\includegraphics{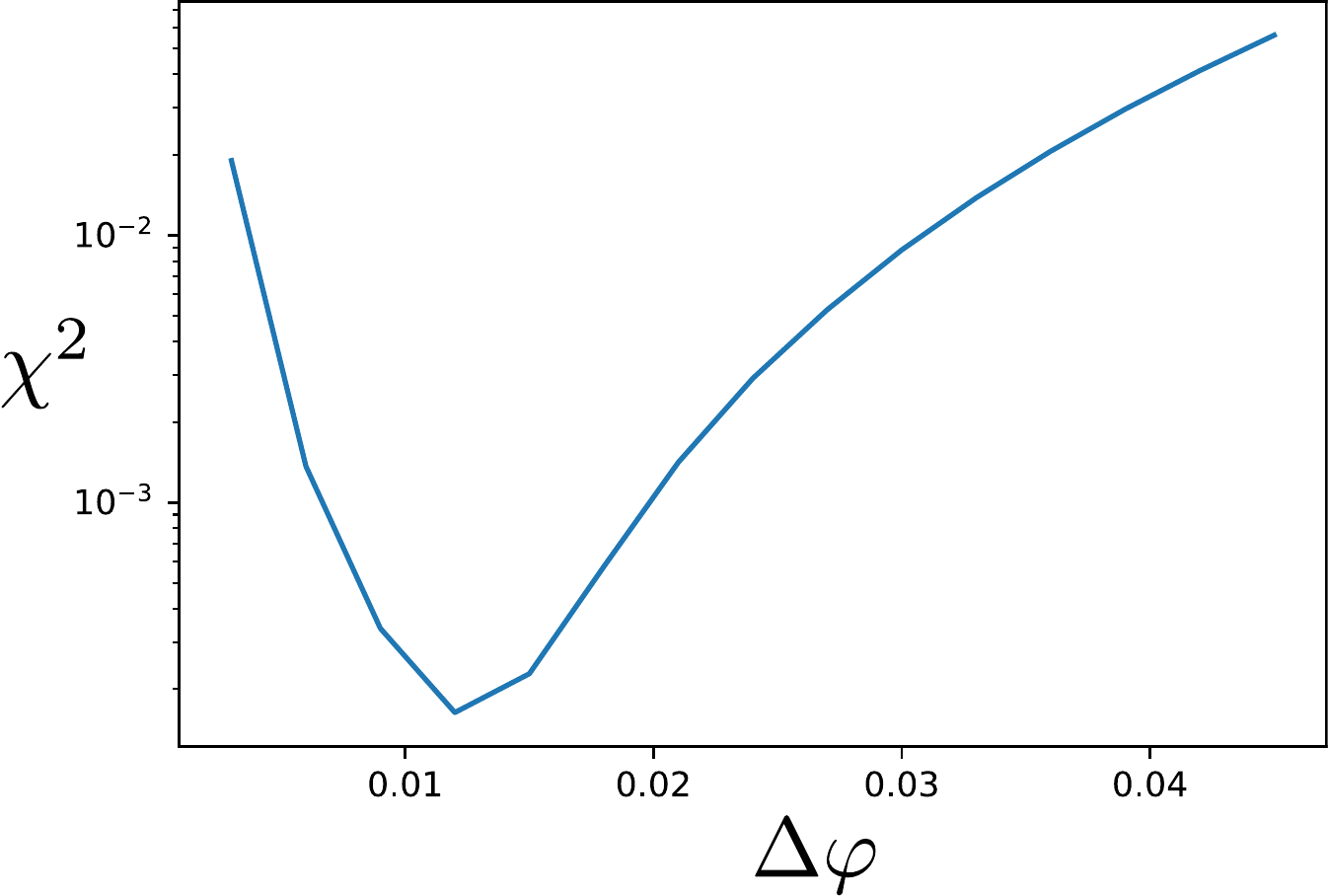}}
\caption{The $\chi^2$ showing the goodness-of-fit between the finite difference derivative and the polynomial derivative.}
    \label{fig:TTFin_Chi}
\end{figure}

In order to calculate the second derivative of the power spectrum we use a finite difference approximation for the second derivative 
\begin{equation} \frac{\partial^2 C_\ell}{\partial \varphi^2}\bigg|_{\varphi=0} \approx \frac{C_\ell(0+\Delta \varphi)-2C_\ell(0)+C_\ell(0-\Delta \varphi)}{(\Delta \varphi)^2}.
\end{equation}
To use this approximation, we must find the step size, $\Delta \varphi$, that gives the most accurate derivative.  To do this, we first fit a polynomial to the power spectrum as a function of $\varphi$, at each multipole moment $\ell$. Fig.~\ref{fig:TTFin_Chi} shows the $\chi^2$ of the polynomial fit, with respect to the actual power spectrum. Note that at small values of $\Delta \varphi$ the $\chi^2$ increases due to residual numerical noise in the derivative. 

Then, we compute the finite difference second derivative for a range of step sizes $\Delta \varphi$ and compare it to the second derivative computed from the polynomial fit. From this procedure we identify a minimum in $\chi^2$ between the finite difference and polynomial second derivatives. The $\chi^2$ between these two methods for the TT spectrum is shown in Fig.~\ref{fig:TTFin_Chi}, and is given by
\begin{equation}
\chi^2 \equiv \frac{\sum_l (y_l - f_l)^2}{\sum_l (y_l)^2},
\end{equation}
where $f_l$ are the finite difference second derivatives of $C_l^{\rm TT}$ and 
$y_l$ are the polynomial fit second derivatives.    We used a step-size of $\Delta \varphi =0.01$ and confirmed that this same step-size allows us to accurately compute the second order derivative of the EE and TE power spectra.

The effects computed in this paper rely on the power spectra between temperature and E-mode polarization and the derivatives of those with respect to $\varphi$.  We show these derivatives in Fig.~\ref{fig:Clderivs}. 

In this work, derivative power spectra $C_{l}^{\rm X,dX'}$ are computed with a suitably modified version of \textsc{Camb} (and \textsc{HyRec} \cite{HyRec}), in which finite-difference derivatives of the CMB transfer functions $X'_{l}(k)$ such that relative convergence relative convergence is exhibited at the $1-10\%$ level.

We show the derivative power spectra in Fig.~\ref{fig:Clderivs}; they are consistent with the numerical derivatives shown in Refs. \cite{Sigurdson:2003pd,OBryan:2013nip}, up to changes in fiducial values of cosmological parameters. Since our constraints (as well as the sensitivity of futuristic experiments) are at the $\sigma_{\alpha}/\alpha_{0}\sim 10^{-3}$ level, the fractional error in CMB two point observables ($\sim 10^{-5}$) is well below cosmic variance  ($\lesssim 10^{-3}$ for the scales of interest), these numerical derivatives are sufficiently accurate for our purpose. 

\end{appendix}

\bibliography{alpha.bib}

\end{document}